\documentclass[10pt,journal]{IEEEtran}

\usepackage{amsmath}
\usepackage{changepage}
\usepackage{algorithm}
\usepackage{algpseudocode}
\usepackage{graphicx}
\usepackage{subcaption}
\usepackage{booktabs}
\usepackage{multirow}
\usepackage{enumitem}
\usepackage{bm}
\usepackage{xcolor}

\usepackage{cite}
\usepackage{balance}

\begin{document}

\title{Reversible Data Hiding over Encrypted Images via Intrinsic Correlation in Block-Based Secret Sharing}

\author{Jianhui~Zou,
        Weijia~Cao,
         Shuang~Yi,
         Yifeng~Zheng,
         and Zhongyun~Hua
%\thanks{This work was supported in part by the National Natural Science Foundation of China under Grants 62071142 and 62201094, and the Guangdong Basic and Applied Basic Research Foundation under Grant 2021A1515011406.}
\thanks{Jianhui~Zou and Zhongyun~Hua are with the School of Computer Science and Technology, Harbin Institute of Technology, Shenzhen, Guangdong 518055, China (E-mail: 23S151139@stu.hit.edu.cn; huazhongyun@hit.edu.cn).}
\thanks{Weijia~Cao is with the Aerospace Information Research Institute, Chinese Academy of Sciences, Beijing 100094, China. (e-mail: caowj@aircas.ac.cn).}
\thanks{Shuang Yi is with the Engineering Research Center of Forensic Science, Chongqing Education Committee, College of Criminal Investigation, Southwest University of Political Science and Law, Chongqing 401120, China. (e-mail: yishuang@swupl.edu.cn).}
\thanks{Yifeng~Zheng is with the Department of Electrical and Electronic Engineering, The Hong Kong Polytechnic University, Hong Kong, China.
 (E-mail: yifengzcs@outlook.com). }
\thanks{This work has been submitted to the IEEE for possible publication. Copyright may be transferred without notice, after which this version may no longer be accessible.}
%\thanks{Yicong~Zhou is with Department of Computer and Information Science, University of Macau, Macau, China (e-mail: yicongzhou@umac.mo).}
}

\IEEEtitleabstractindextext{
\begin{abstract}
With the rapid advancements in information technology, reversible data hiding over encrypted images (RDH-EI) has become essential for secure image management in cloud services. However, existing RDH-EI schemes often suffer from high computational complexity, low embedding rates, and excessive data expansion. This paper addresses these challenges by first analyzing the block-based secret sharing in existing schemes, revealing significant data redundancy within image blocks. Based on this observation, we propose two space-preserving methods: the direct space-vacating method and the image-shrinking-based space-vacating method. Using these techniques, we design two novel RDH-EI schemes: a high-capacity RDH-EI scheme and a size-reduced RDH-EI scheme. The high-capacity RDH-EI scheme directly creates embedding space in encrypted images, eliminating the need for complex space-vacating operations and achieving higher and more stable embedding rates. In contrast, the size-reduced RDH-EI scheme minimizes data expansion by discarding unnecessary shares, resulting in smaller encrypted images. Experimental results show that the high-capacity RDH-EI scheme outperforms existing methods in terms of embedding capacity, while the size-reduced RDH-EI scheme excels in minimizing data expansion. Both schemes provide effective solutions to the challenges in RDH-EI, offering promising applications in fields such as medical imaging and cloud storage.
\end{abstract}

% Note that keywords are not normally used for peerreview papers.
\begin{IEEEkeywords}
Secret sharing, reversible data hiding, encrypted images.
\end{IEEEkeywords}}

% make the title area
\maketitle

\IEEEdisplaynontitleabstractindextext

\IEEEpeerreviewmaketitle

\ifCLASSOPTIONcompsoc
\IEEEraisesectionheading{\section{Introduction}\label{sec:introduction}}
\else
\section{Introduction}\label{sec:introduction}
\fi

\IEEEPARstart{A}{s} information technology advances rapidly, the volume of images generated daily by a diverse range of imaging devices is growing at an unprecedented rate. Cloud services have become essential for managing this growing influx of images, offering benefits such as seamless access, secure backups, and scale storage capacity~\cite{wang2018three}. However, to manage the store images, cloud servers often embed metadata and authentication information, potentially altering the original content~\cite{zhou2015secure,wang2021reversible,qian2016separable}. Such modifications are unacceptable in sensitive fields like healthcare, law, and scientific research. To address this, reversible data hiding (RDH) techniques have been developed, utilizing approaches such as lossless compression~\cite{celik2005lossless}, difference expansion~\cite{tian2003reversible}, histogram shifting~\cite{gao2011lossless,jia2019reversible}, prediction error expansion~\cite{thodi2007expansion,qi2023reversible,yuan2024reversible}, and deep learning methods~\cite{hu2021reversible, chang2022reversible,zhou2023reversible}. Furthermore, to ensure data privacy, users often opt to encrypt images before uploading them to the cloud. This has led to growing interest in reversible data hiding over encrypted images (RDH-EI), a method that enables cloud servers to embed management data into encrypted images while maintaining user privacy.

In RDH-EI systems, three main participants are involved: the content owner, the data hider, and the receiver. The process begins with the content owner encrypting the images before uploading them to the cloud. The cloud server, acting as the data hider, embeds the necessary management data into these encrypted images. Depending on the keys available, the receiver can either extract the hidden data or fully recover the original image, achieving both data embedding and image restoration in a single step. This framework ensures that the original image remains intact while allowing for the embedding of additional data.

RDH-EI schemes can be categorized based on the timing of space allocation for data embedding: those that reserve room before encryption (RRBE)\cite{zhang2024reversible, ji2023reversible, liu2024bi, yao2024reversible} and those that vacate room after encryption (VRAE)\cite{xiao2024high, zheng2019lossless, ke2021reversible, hua_reversible_2023, yu_reversible_2023}. RRBE schemes involve the content owner pre-allocating space for embedding data before encryption, resulting in a fixed embedding capacity. These schemes tend to offer higher embedding capacity than VRAE-based methods, as the content owner can leverage strong pixel correlations inherent in natural images~\cite{yin2020reversible, yu2021reversible, qiu_high-capacity_2022}. However, the need to pre-reserve space for embedding adds significant computational overhead. On the other hand, VRAE-based schemes often employ stream ciphers, block-based encryption, or homomorphic encryption. Early methods that used stream ciphers disrupted pixel correlation, resulting in low embedding capacity~\cite{zhang2011reversible, zhang2011separable, wu2014high, liao2015reversible, qian2016encoding}. To improve capacity, more recent schemes have used block-based encryption techniques, such as block-based permutation~\cite{huang2016new, qiu_high-capacity_2022} and co-XOR~\cite{gao2022high}, which retain partial pixel correlation within blocks, allowing for higher embedding capacity. However, these methods are vulnerable to known-plaintext attacks~\cite{qu2021cryptanalysis}. Homomorphic encryption provides stronger security by allowing computations on encrypted data~\cite{ke_fully_2020, li_histogram_2017, zheng2019lossless, wu2023homomorphic}, but it is computationally expensive and results in data expansion.

Among these methods, secret sharing-based schemes are recently attractive due to their ability to resist single-point failures. In these schemes, space for embedding data is typically vacated using block-based secret sharing. While effective, these space-vacating techniques are computationally complex and highly reliant on pixel correlation. When pixel correlation is weak in the original image, embedding data becomes challenging. This issue arises from the inherent trade-offs in block-based secret sharing, where identical random numbers are used within each block, leading to intrinsic correlation. Previous schemes often overlook this intrinsic correlation, focusing solely on the pixel correlation generated during the encryption process. To enhance performance while maintaining security, we investigate the correlation within the encryption process itself. Our findings reveal that block-based secret sharing utilizes fewer random numbers within each block, thereby reducing the number of pixel shares required for recovery and improving data embedding efficiency. Based on these insights, we propose two novel space-vacating methods and introduce two RDH-EI schemes: one optimized for achieving high data embedding capacity, and the other focused on reducing the size of the encrypted images. The key contributions and innovations of this paper are as follows:

\begin{enumerate}
\item We identify an important intrinsic correlation in block-based secret sharing, where identical random numbers are used within each block, thereby reducing the number of pixel shares required for recovery. 

\item Building on this intrinsic correlation, we propose two space-vacating techniques for data embedding: the direct space-vacating method and the image-shrinking-based space-vacating method.

\item Using the two data embedding techniques, we further design two RDH-EI schemes: a high-capacity RDH-EI scheme and
a size-reduced RDH-EI scheme. The high-capacity RDH-EI
scheme aims to achieve high data embedding capacity, while the size-reduced RDH-EI scheme is designed to reduce the size of the encrypted images. 

\item Experimental results demonstrate the high embedding
capacity of the high-capacity RDH-EI scheme and the low data expansion of the size-reduced RDH-EI scheme. 
\end{enumerate}

The paper is organized as follows. Section~\ref{sec:related work} reviews existing secret sharing-based RDH-EI schemes. Section~\ref{sec:the proposed space-vacating methods} introduces the intrinsic correlation when secret sharing is performed based on blocks and proposes two novel space-vacating methods. Section~\ref{sec: scheme I} and Section~\ref{sec:scheme II} present a scheme for direct embedding space generation and a scheme with reduced data expansion. Section~\ref{sec: Experimental Results} provides simulation results, and Section~\ref{sec: Comparison and Analysis} compares our schemes with state-of-the-art methods. Finally, Section~\ref{sec: conclusion} concludes the paper.

\section{RELATED WORK} \label{sec:related work}
Recently, some RDH-EI schemes using secret sharing techniques have been proposed~\cite{wu2018adopting, chen2019new, chen_secret_2020, ke2021reversible, qin2021reversible, hua2022matrix, hua_reversible_2023, yu_reversible_2023}. These schemes provide the ability to resist single-point failures over traditional encryption techniques. In $(r,n)$ secret sharing, the original pixel is encrypted into $n$ shares, and any $r$ ($r\leq n$) of those shares are sufficient to recover the original pixel value. By distributing the image shares on different cloud servers, these schemes ensure robustness against data loss or corruption. Based on the way to perform secret sharing, these schemes can be classified into two categories: block-based schemes and non-block-based schemes.

\subsection{Block-Based Schemes}
Block-based secret sharing schemes partition images into smaller blocks, creating space for data embedding through space-vacating techniques similar to those in block-based lightweight encryption. These techniques typically involve predicting pixel values using a pixel predictor, calculating prediction errors,  and then applying compression methods to minimize these errors. The primary distinctions between different block-based secret sharing schemes lie in the size of the blocks, the choice of pixel predictors, and the compression methods used. In Chen$~et~al.$'s scheme~\cite{chen2019new}, each block consists of a pixel pair, and the predictor computes the difference between the two pixel values in each pair. Various expansion techniques are used for compression.  Qin$~et~al.$~\cite{qin2021reversible} use $2 \times 2$ blocks and introduce a novel compression method based on preserving differences. Here, prediction errors determine the embedding capacity, with some pixel bits representing these errors while others are used for data embedding. The scheme in ~\cite{ke2021reversible} is similar to that of Chen$~et~al.$, but incorporates the Chinese remainder theorem for secret sharing. In~\cite{hua2022matrix,hua_reversible_2023}, block sizes of $2 \times 2$, $4 \times 4$, or $8 \times 8$ are used, with the median edge detector (MED) employed as the pixel predictor. The authors propose two compression methods that leverage arithmetic coding and block properties to enhance embedding capacity, along with two novel secret-sharing techniques based on matrix theory. Yu$~et~al.$~\cite{yu_reversible_2023} use block sizes of $4 \times 4$ or $8 \times 8$ and employ the MED for pixel prediction. They propose a hybrid coding method where each block dynamically selects between hierarchical coding and arithmetic coding to optimize embedding rates. While some of these schemes achieve high embedding capacity, their space-vacating methods are computationally expensive. Additionally, the embedding rate is highly dependent on the pixel correlation of the image. When pixel correlation is low, it becomes difficult to create sufficient space for embedding, reducing the effectiveness of block-based schemes. Furthermore, these approaches typically result in high data expansion rates, which can negatively impact transmission and storage efficiency. An exception is Chen~\textit{et al.}'s scheme~\cite{chen2019new}, which, although efficient, offers limited security.

\subsection{Non-Block-Based Schemes}
Chen \textit{et al.}~\cite{chen_secret_2020} proposed a scheme that directly generates the embedding space through secret sharing, bypassing pixel correlation. As a result, the secret sharing is performed based on pixels. Given an image of size $M \times N$, the content owner preprocesses it and uses an encryption key to generate $M \times N$ pseudorandom numbers. Each pseudorandom number becomes the first polynomial coefficient, with each original pixel serving as the second coefficient, resulting in $M \times N$ secret-sharing polynomials. For each polynomial, $n$ shares are generated, one of which is designated for data embedding. During decryption, the receiver regenerates the pseudorandom numbers using the encryption key. If a corrupted share is present, the receiver iterates through possible values for the damaged share. Lagrangian interpolation verifies if the recovered first polynomial coefficient matches the pseudorandom number, identifying the second coefficient as the original pixel value. The embedding rate of this scheme is $l/n$, where $n$ is the number of encrypted images and $l$ is the number of bit-planes replaced in the disrupted shares. The maximum embedding rate is 3.5 \textit{bpp}. While it overcomes pixel correlation limitations, the scheme has drawbacks: the embedding rate is limited, the data expansion rate is $n$, and the encryption process requires preprocessing. Additionally, decryption involves computationally expensive exhaustive search techniques.

\subsection{Discussions}
Secret sharing-based RDH-EI schemes can resist single-point failures and offer high embedding rates, making them more suitable in practical applications compared to traditional encryption-based schemes. However, existing schemes still have some drawbacks. In block-based schemes, the embedding rate depends on pixel correlation, and the space-vacating process is complex. In non-block-based schemes, the embedding rate is limited, and the decryption process is time-intensive. Both approaches also suffer from data expansion issues. To address these challenges, we propose novel space-vacating methods for block-based secret sharing and introduce two new block-based secret sharing schemes that effectively address these challenges.
%we recognize that high correlation exists not only within encrypted pixel blocks but also within the secret sharing method itself. Specifically, fewer random numbers are used during encryption, fewer shares are needed for decryption. Leveraging this correlation, we propose a novel decryption method for block-based secret sharing and introduce two block-based secret sharing schemes that address these problems effectively.

\section{The Proposed Space-Vacating Methods}\label{sec:the proposed space-vacating methods}
In this section, we first introduce the encryption process of block-based secret sharing which is always used in previous schemes. There is a correlation in the encryption phase when secret sharing is performed based on blocks. We discuss it and then propose two novel space-vacating methods. 
\subsection{The Process of Block-Based Secret Sharing}\label{sec: encryption}
To generate space for data embedding, the original image is divided into blocks and secret sharing is performed based on image blocks. In this way, part of pixel correlation is reserved in blocks.

For grayscale images, pixels fall within the range \([0, 256)\), making \(GF(2^8)\) an ideal field for secret sharing. During the encryption stage, the image \(\bm{I}\) with dimensions \(M \times N\) is encrypted block-by-block using Shamir's \((r, n)\) secret sharing~\cite{shamir_how_1979}. Let the block size be \(S\). For the \(i\)-th image block \(\bm{B}_i\) (\(i \in \{0, 1, \dots, BN-1\}\), where \(BN = \frac{M \times N}{S\times S}\)), the content owner generates \(r-1\) random integers \(a(0), a(1), \dots, a(r-2)\) within the range \([0, 256)\). Let \(p(y) = y^8 + y^4 + y^3 + y + 1\) be an irreducible polynomial over a finite field. The following polynomial is constructed for modulo \(p(y)\):
\begin{equation}
\begin{aligned}
    F_i(j, x) =  \bm{B}_i(j) \oplus a(0) x &\oplus \cdots \oplus \\ & a(r-2) x^{r-1} \mod p(y),
\end{aligned}
\label{eq:used Shamir}
\end{equation}
where \(\bm{B}_i(j)\) is the \(j\)-th pixel in \(\bm{B}_i\). Using the encryption key \(K_E\), the content owner generates \(n\) distinct nonzero integers \(x(0), x(1), \dots, x(n-1)\) within the range \([0, 2^8)\). Then, the content owner obtains the encrypted pixels \(F_i(j, x(0)), F_i(j, x(1)), \dots, F_i(j, x(n-1))\), corresponding to the \(n\) encrypted pixel in \(n\) encrypted images.

After all image blocks are encrypted, \(n\) encrypted images are produced. Note that the parameters \(a(0), a(1), \dots, a(r-2)\), \(x(0), x(1), \dots, x(n-1)\) remain fixed for encrypting pixels within each block, but they vary across blocks.

\subsection{The Intrinsic Correlation in Block-Based Secret Sharing}\label{sec:the correlation}
Block-based secret sharing uses only \(r-1\) random numbers \(a(0), a(1), \dots, a(r-2)\) for encrypting each original block. The reused random numbers introduce correlation in the encryption process.

We represent all secret-sharing polynomials as a system of equations. It is a system of nonlinear equations. Let \(BS\) be equal to \(S^2\), which represents the number of pixels within the block. There are only \( BS+r-1\) unknowns in the system of equations but the number of equations is $BS\times n$. There are too many equations in this set. Since the equations correspond directly to the unknowns, it means that certain equations are unnecessary for solving the unknowns.
We try to find a system consisting of \(BS+r-1\) equations to solve the unknowns. Denote \(p_c\) is the chosen pixel in \(\bm{B}_i\). The system consists of any \(r\) equations of \(p_c\), and any one equation of the remaining pixels in \(\bm{B}_i\). It is a nonhomogeneous system of linear equations. Let \(\bm{X}\) represents the matrix composed of \(x\),\(\bm{a}\) represents the column vector composed of \(a\) and \(p\), and \(\bm{f}\) represents the vector composed of the obtained share values, \(c(0),c(1),\cdots,c(r-1)\) represent \(r\) distinct integers in \([0,n)\) using for \(p_c\), \(l(0), \cdots,l(c-1),l(c+1),\cdots,l(BS-1)\) represent \(BS-1\) integers in \([0,n)\). The system of equations can be described as 
\begin{equation}
    \bm{f} = \bm{X \oplus a} \mod p(y),
\end{equation}
where \(\bm{f} = 
     \left[f(c,c(0)), \cdots, f(c,c(r-1)), f(1,l(0)), \cdots, f(c - 1, \right. \\\left. l(c - 1)),  f(c + 1,l(c + 1)), \cdots, f(BS-1,l(BS-1))\right]^\mathrm{T}\),
\(\bm{a} = \left[\mathbf{IB}(0), \cdots,  \mathbf{IB}(c - 1), \mathbf{IB}(c), 
  \mathbf{IB}(c + 1), \cdots, \mathbf{IB}(BS-1), \right. \\\left. a(0),  \cdots, a(r - 2)\right]^\mathrm{T}
\)
, and \(\bm{X}=\)
\begin{equation}
\footnotesize
\begin{bmatrix}
0 & \cdots & 0 & 1 & 0 & \cdots & 0 & x(c(0)) & \cdots & x(c(0))^{r-1} \\
\vdots & \ddots & \vdots & \vdots & \vdots & \ddots & \vdots & \vdots & \ddots & \vdots \\
0 & \cdots & 0 & 1 & 0 & \cdots & 0 & x(c(r-1)) & \cdots & x(c(r-1))^{r-1} \\
1 & \cdots & 0 & 0 & 0 & \cdots & 0 & x(l(0)) & \cdots & x(l(0))^{r-1} \\
\vdots & \ddots & \vdots & \vdots & \vdots & \ddots & \vdots & \vdots & \ddots & \vdots \\
0 & \cdots & 1 & 0 & 0 & \cdots & 0 & x(l(c-1)) & \cdots & x(l(c-1))^{r-1} \\
0 & \cdots & 0 & 0 & 1 & \cdots & 0 & x(l(c+1)) & \cdots & x(l(c+1))^{r-1} \\
\vdots & \ddots & \vdots & \vdots & \vdots & \ddots & \vdots  & \vdots & \ddots & \vdots \\
0 & \cdots & 0 & 0 & 0 & \cdots & 1 & x(l(BS-1)) & \cdots & x(l(BS-1))^{r-1} \\
\end{bmatrix}
\end{equation}

% \begin{equation*}
% \small
% 	\label{eqa.kafc}
% 	 \begin{bmatrix}
% 		f(c,1)\\
% 		\vdots\\
% 		f(c,r)\\
%             f(1,l(1))\\
%             \vdots\\
%             f(c-1,l(c-1))\\
%             f(c+1,l(c+1))\\
%             \vdots\\
%             f(BS,l(BS))
% 	\end{bmatrix} =
% 	\begin{bmatrix}
% 0 & \cdots & 0 & 1 & 0 & \cdots & 0 & x(c(1)) & \cdots & x(c(1))^r \\
% \vdots & \ddots & \vdots & \vdots & \vdots & \ddots & \vdots & \vdots & \ddots & \vdots \\
% 0 & \cdots & 0 & 1 & 0 & \cdots & 0 & x(c(r)) & \cdots & x(c(r))^r \\
% 1 & \cdots & 0 & 0 & 0 & \cdots & 0 & x(l(1)) & \cdots & x(l(1))^r \\
% \vdots & \ddots & \vdots & \vdots & \vdots & \ddots & \vdots & \vdots & \ddots & \vdots \\
% 0 & \cdots & 1 & 0 & 0 & \cdots & 0 & x(l(c-1)) & \cdots & x(l(c-1))^r \\
% 0 & \cdots & 0 & 0 & 1 & \cdots & 0 & x(l(c+1)) & \cdots & x(l(c+1))^r \\
% \vdots & \ddots & \vdots & \vdots & \vdots & \ddots & \vdots  & \vdots & \ddots & \vdots \\
% 0 & \cdots & 0 & 0 & 0 & \cdots & 1 & x(l(BS)) & \cdots & x(l(BS))^r \\
        
% 	\end{bmatrix}\times \begin{bmatrix}
% 	\mathbf{IB}(1) \\
%         \vdots \\
%         \mathbf{IB}(c-1)\\
%         \mathbf{IB}(c)\\
%         \mathbf{IB}(c+1)\\
%         \vdots\\
%         \mathbf{IB}(BS)\\
% 	a_1\\
% 	\vdots\\
% 	a_{r-1}\\
% 	\end{bmatrix} \mod p(y),
% \end{equation*}

To prove that this system of equations has a unique solution, we need to demonstrate that the determinant of \(\bm{X}\) is non-zero, according to Cramer's Rule~\cite{robinson1970short}. The proof proceeds as follows:

First, we perform successive expansions on the rows of the matrix, starting from the last \(BS - 1\) rows. For the \((r+1)\)-th row, we expand it and obtain that \(\det(\bm{X}) = \)
\[
\footnotesize
\begin{vmatrix}
0 & \cdots & 0 & 1 & 0 & \cdots & 0 & x(c(0)) & \cdots & x(c(0))^{r-1} \\
\vdots & \ddots & \vdots & \vdots & \vdots & \ddots & \vdots & \vdots & \ddots & \vdots \\
0 & \cdots & 0 & 1 & 0 & \cdots & 0 & x(c(r-1)) & \cdots & x(c(r-1))^{r-1} \\
1 & \cdots & 0 & 0 & 0 & \cdots & 0 & x(l(0)) & \cdots & x(l(0))^{r-1} \\
\vdots & \ddots & \vdots & \vdots & \vdots & \ddots & \vdots & \vdots & \ddots & \vdots \\
0 & \cdots & 1 & 0 & 0 & \cdots & 0 & x(l(c-1)) & \cdots & x(l(c-1))^{r-1} \\
0 & \cdots & 0 & 0 & 1 & \cdots & 0 & x(l(c+1)) & \cdots & x(l(c+1))^{r-1} \\
\vdots & \ddots & \vdots & \vdots & \vdots & \ddots & \vdots  & \vdots & \ddots & \vdots \\
0 & \cdots & 0 & 0 & 0 & \cdots & 1 & x(l(BS-1)) & \cdots & x(l(BS-1))^{r-1} \\
\end{vmatrix}
\]
Next, we repeat this row expansion process for the remaining \(BS - 2\) rows. After completing the operation, we obtain the following determinant expression:
\[
\det(\bm{X}) = 
\begin{vmatrix}
1 & x(c(0)) & \cdots & x(c(0))^{r-1} \\
\vdots & \vdots & \ddots & \vdots \\
1 & x(c(r-1)) & \cdots & x(c(r-1))^{r-1}
\end{vmatrix}
\label{eq:detX}
\]
This determinant is a Vandermonde determinant, and its value is given by the well-known formula:
\[
\det(\bm{X}) = \prod_{0 \leq j < i \leq r-1} \left( x(c(i)) \oplus x(c(j)) \right)
\]
as shown in~\cite{horn2013matrix}. Since all the \(x\)s are distinct, \(\det(\bm{X}) \neq 0\). Therefore, this system of equations has a unique solution.

Based on the proof, we know that it can recover the original block with only \(BS+r-1\) equations. In the decryption phase, we recover the original image as follows:

\begin{enumerate}
    \item For the original block \(\bm{B}_i\), randomly choose a pixel \(\hat{p}\).
    \item Collect any \(r\) shares of \(\hat{p}\), then use Lagrangian interpolation and \(K_E\) to recover \(\hat{p}\) and the random numbers \(a(0), a(1), \dots, a(r-2)\).
    \item For the remaining pixels in \(\bm{B}_i\), collect any one share \(F_i(j, x(l(j))\) for each pixel. Then, use the following equation:
    \begin{equation}
    \begin{aligned}
        \bm{B}_i(j) = & F_i(j, x(l(j)) \oplus a(0) x(l(j)) \\ & \oplus \cdots \oplus a(r-2) x(l(j))^{r-1} \mod p(y),
    \label{eq:decrypt Shamir}
    \end{aligned}
    \end{equation}
    where \(x(l)\) is generated by \(K_E\).
    \item Repeat steps 1)-3) for each block in \(\bm{I}\).
\end{enumerate}
To derive Eq.~(\ref{eq:decrypt Shamir}), we start from Eq.~(\ref{eq:used Shamir}). After performing an XOR operation with \(a(0)x \oplus \cdots \oplus a(r-2)x^{r-1}\) on both sides of Eq.~(\ref{eq:used Shamir}), we reduce the result modulo \(p(y)\) over \(GF(2^8)\), leading to the conclusion:
\begin{align*}
    F_i(j, x) \oplus  &a(0) x \oplus \cdots \oplus \\& a(r-2) x^{r-1} \mod p(y) = \bm{B}_i(j) \mod p(y).
\end{align*}
Then, by transposing the equation and setting \(x = x(l(j))\), we obtain Eq.~(\ref{eq:decrypt Shamir}).

\subsection{The Direct Space-Vacating Method}\label{sec: the direct space-vacating method}

As described in Section~\ref{sec:the correlation}, for recovery in each image block, \(r\) shares of \(\hat{p}\) and \(BS - 1\) shares of the remaining pixels are required. However, in order to ensure that any \(r\) encrypted images can reconstruct the original image, additional shares must be retained. This section outlines the specific share retention requirements and how the embedding space is organized for data hiding.

\subsubsection{Shares for \(\hat{p}\)}

We first focus on \(\hat{p}\), which is the key pixel used in the recovery process. Only \(r\) shares of its are needed for recovery. If one of \(n\) shares is damaged, the collected \(r\) shares may include the damaged share, preventing recovery. Therefore, in order to ensure robustness in the recovery process, all \(n\) shares of \(\hat{p}\) must be retained. This ensures that no single damaged share can disrupt the recovery process.

\subsubsection{Shares for Other Pixels}

For the remaining pixels, consider the case where \(r\) out of \(n\) shares are damaged. In this case, all the collected \(r\) shares could be damaged, preventing recovery. To mitigate this, if up to \(r-1\) shares are damaged, the collected shares will still contain at least one valid share, allowing the pixel value to be recovered during the decryption phase. Hence, it is necessary to retain \((BS - 1) \times (n - r + 1)\) shares of the other pixels.

\subsubsection{Embedding Space Allocation}

Given the above retention requirements, only the shares corresponding to the remaining pixels in each block can be used for data hiding. These shares must be distributed carefully to ensure the embedding capacity is evenly spread across all encrypted images. Specifically, the embedding space is organized as follows:

\begin{enumerate}
    \item The first pixel in each block is designated as the \(p_c\), and all shares corresponding to this pixel are retained. This means that the first pixel in each encrypted image serves as a key reference point and does not participate in the data embedding process.
    \item For the remaining pixels, we partially assign their shares as embedding space using a sliding window approach. Let \(\bm{EP}\) be the array consisting of all pixels except the first pixels in each block. For a given pixel indexed by \(i\), its specific \(r-1\) shares are used as the embedding space. Specifically, the shares \(\bm{EP}(i \mod n, i), \bm{EP}((i+1) \mod n, i), \dots, \bm{EP}((i+r-2) \mod n, i)\) are allocated as the embedding space. 
\end{enumerate}

This arrangement ensures that the difference in embedding capacity between different encrypted images does not exceed \((n-r+1) \times 8\) bits, thereby maintaining balance across the images.

\subsubsection{Embedding Space Retrieval Algorithm}

To facilitate the embedding of data into the encrypted images, we propose Algorithm~\ref{alg:embedding space}, which retrieves the embedding space within \(\bm{EI}(ID)\) based on its identity \(ID\). This algorithm takes as input the number of image blocks, the block size, the secret sharing threshold, the number of encrypted images, and the encrypted image identity, and outputs an array \(\bm{\overline{EP}}(ID)\) consisting of pixels that can be embedded in \(\bm{EP}\).

\begin{algorithm}[!t]
\caption{\emph{Find Embedding Space by ID}}
\label{alg:embedding space}
\begin{algorithmic}[1]
\renewcommand{\algorithmicrequire}{\textbf{Initialize:}} 
\renewcommand{\algorithmicensure}{\textbf{Output:}}

\Require The number of image blocks \(BN\), block size \(S\), secret sharing threshold \(r\), the number of encrypted images \(n\), encrypted image identity \(ID\), and the array \(\bm{EP}(ID)\) consisting of pixels except the first pixels in blocks of the \(ID\)-th encrypted image.
\Ensure An array \(\bm{\overline{EP}}(ID)\) consisting of pixels that can be embedded in \(\bm{EP}\).

\State Initialize \(Iter = 0\)
\If{$ID \leq r-1$:} 
    \State \(Iter = ID - r + 2\)
    \State \(\bm{\overline{EP}}(ID) = \text{append}(\bm{\overline{EP}}(ID), \bm{EP}(ID, Iter))\)
\EndIf
\For{$Iter < BN \times (BS - 1)$:}
    \State Initialize \(SP = ID - r + 2\)
    \State Initialize \(BIAS = Iter - SP \mod n\)
    \If{$BIAS == r-2$:} 
        \State \(Iter = Iter + n - r + 2\)
    \Else{}
        \State \(Iter = Iter + 1\)
    \EndIf
    \State \(\bm{\overline{EP}}(ID) = \text{append}(\bm{\overline{EP}}(ID), \bm{EP}(ID, Iter))\)
\EndFor
\end{algorithmic}
\end{algorithm}

This algorithm efficiently retrieves the embedding space in the encrypted image, facilitating the data embedding process.

\subsection{Image-Shrinking-Based Space-Vacating Method}\label{sec: image-shrinking-based space-vacating method}

In contrast to the method described in Section~\ref{sec: the direct space-vacating method}, this method focuses on retaining only the pixel shares necessary for recovery while discarding the remaining pixel shares in the encrypted images. This method minimizes storage requirements by eliminating redundant shares while preserving the correlation among some blocks.

\subsubsection{Pixel Share Retention and Discarding Strategy}

For each image block, the chosen pixel \(\hat{p}\) is set as the first pixel, and all shares corresponding to \(\hat{p}\) are retained. This ensures that the key pixel used for image recovery remains intact. For the remaining pixels in the block, shares are discarded based on the index of the block. This approach maximizes the preservation of pixel correlations within the block.

In more detail, for the encrypted blocks \(\bm{EB}_i\) at positions indexed by \((i \mod n)\)-th, \((i+1 \mod n)\)-th, \(\dots\), \((i + r-2 \mod n)\)-th, only their first pixel shares are retained. For the remaining \(\bm{EB}_i\) blocks, all pixel shares are retained. This selective share retention strategy ensures that pixel correlation is preserved in blocks while unnecessary data is discarded.

\subsubsection{Prediction and Error Calculation}

After applying the pixel share retention and discarding strategy, the next step is to calculate the prediction errors. A predictor is used to generate prediction values for the pixels, and the difference between the predicted and actual pixel values is considered the prediction error.

\begin{figure}[!t]
\centering
\begin{subfigure}[b]{0.10\textwidth}
\includegraphics[width=\linewidth]{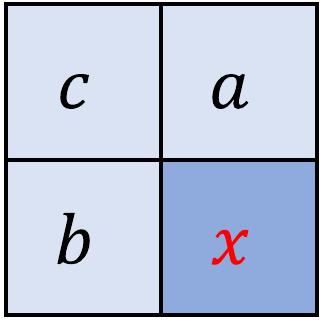}
\end{subfigure}
\caption{Shape of the MED predictor, where \(x\) is the value to be predicted.}
\label{Fig:predictors}
\end{figure}

In this method, the MED predictor~\cite{weinberger2000loco}, illustrated in Fig.~\ref{Fig:predictors}, is employed to calculate the predicted values for the pixels in the encrypted blocks.

For a given pixel \(p(j,k)\) located at position \((j,k)\) in the encrypted block \(\bm{EB}\), its predicted value \(\hat{p}(j,k)\) is computed using the MED predictor as follows:
\begin{equation}
\label{eq.med}
\begin{aligned}
\hat{p}(j,k)=
\begin{cases}
\max(a,b) & \text{if } c \leq \min(a,b),\\
\min(a,b) & \text{if } c \geq \max(a,b),\\
a + b - c &\text{otherwise},
\end{cases}
\end{aligned}
\end{equation}
where \(a = p(j-1,k)\), \(b = p(j,k-1)\), and \(c = p(j-1,k-1)\) are the neighboring pixels used for prediction.

When applying the MED predictor, edge pixels (located at the first row and the first column) present a challenge, as they do not have all the required neighbors for prediction. To handle this, we set \(a = b = c\) for these edge pixels, as shown in Fig.~\ref{Fig:predictors}, allowing the prediction to proceed effectively.

After calculating the predicted value \(\hat{p}(j,k)\), the prediction error \(e(j,k)\) is computed as:
\begin{equation}
e(j,k) = p(j,k) - \hat{p}(j,k),
\end{equation}
where \(e(j,k)\) represents the prediction error at pixel \((j,k)\) in the error block \(\bm{ERB}\).

\subsubsection{Error Compression}

Once the prediction errors have been calculated for all pixels in the encrypted blocks, the next step is to compress these errors efficiently. Arithmetic coding is used for the compression of the error blocks. The compressed bitstream \(CB\) is generated by applying arithmetic coding to \(\bm{ERB}\).

To record side information \(SI\) for the compressed bitstream, we proceed as follows:
\begin{itemize}
    \item The length of \(CB\) is recorded in \(SI\) using \(\log(8 \times M' \times N')\) bits, where \(M'\) and \(N'\) represent the size of the compressed image, which is smaller than the original size \(M \times N\).
    \item The counts of each error value are also recorded in \(SI\) using \(511 \times \log(M' \times N')\) bits for decoding.
\end{itemize}
Finally, both the side information \(SI\) and the compressed bitstream \(CB\) are embedded into the pixels of the encrypted blocks, excluding the first pixel in each block. The remaining pixels serve as the embedding space for this data.

\section{The Proposed High-Capacity RDH-EI Scheme} \label{sec: scheme I}
Based on Section~\ref{sec: the direct space-vacating method}, we present an RDH-EI scheme that directly generates embedding space. The proposed scheme allows the data hider to embed data without using pixel correlation.

\subsection{Content Owner}
The content owner encrypts the original image using the encryption key \(K_E\) to generate \(n\) encrypted images, as described in Section~\ref{sec: encryption}. The image is divided into non-overlapping blocks of size \(BS\) and then encrypted using \((r, n)\)-threshold Shamir's secret sharing to produce \(n\) image shares.

Additionally, the parameters \(S\), \(r\), \(n\), and \(ID\) are embedded into the encrypted images, where \(S\) represents the block size, \(r\) denotes the threshold of Shamir's secret sharing, \(n\) is the number of generated shares, and \(ID\) indicates the identity of the encrypted image. Specifically, the first pixel in each of the first four blocks is replaced with \(S\), \(r\), \(n\), and \(ID\), respectively. 

Using Algorithm~\ref{alg:embedding space} from Section~\ref{sec: the direct space-vacating method}, the content owner determines the embedding space of each encrypted image. The four original pixel values from the first four blocks are embedded into \(\bm{\overline{EP}}(ID, 0)\), \(\bm{\overline{EP}}(ID, 1)\), \(\bm{\overline{EP}}(ID, 2)\), and \(\bm{\overline{EP}}(ID, 3)\). Finally, the \(n\) encrypted images are generated and sent to \(n\) different data hiders.

\subsection{Data Hider}
Upon receiving an encrypted image \(\bm{EI}(\beta)\),  the \(\beta\)-th data hider can independently embed a secret message into it, generating the \(\beta\)-th marked encrypted image \(\bm{MI}(\beta)\). First, the data hider extracts the first pixel to obtain \(S\). The encrypted image \(\bm{EI}(\beta)\) is then divided into non-overlapping blocks of size \(S\times S\). Next, the parameters \(r\), \(n\), and \(ID\) are extracted from the first pixels of the second, third, and fourth blocks. The number of blocks \(BN\) can be calculated as \(BN = \frac{M \times N}{S\times S}\).

With the parameters \(BN\), \(S\), \(r\), \(n\), \(ID\), and \(\bm{EP}\), the data hider uses Algorithm~\ref{alg:embedding space} to determine the embedding space \(\bm{\overline{EP}}(ID)\). The data hider then embeds the secret message into these identified spaces, excluding \(\bm{\overline{EP}}(ID, 0)\), \(\bm{\overline{EP}}(ID, 1)\), \(\bm{\overline{EP}}(ID, 2)\), and \(\bm{\overline{EP}}(ID, 3)\). To enhance security, the embedded data can be encrypted using a cryptographic algorithm (e.g., AES) with a data hiding key \(K_D\).

\subsection{Receiver}
A receiver with the encryption key \(K_E\) and \(r\) marked encrypted images can reconstruct the original image. Additionally, a receiver with the data hiding key \(K_D\) can extract the embedded data from the marked encrypted images.

\subsubsection{Decryption}
The receiver obtains the \(r\) marked encrypted images \(\bm{MI}(\gamma_0), \bm{MI}(\gamma_1), \dots, \bm{MI}(\gamma_{r-1})\). For each marked encrypted image, the receiver extracts the first pixel from the first four blocks to obtain \(S\), \(r\), \(n\), and \(ID\), and then extracts \(\bm{\overline{EP}}(ID, 0)\), \(\bm{\overline{EP}}(ID, 1)\), \(\bm{\overline{EP}}(ID, 2)\), and \(\bm{\overline{EP}}(ID, 3)\), placing them back into the first pixels of the first four blocks of \(\bm{MI}(ID)\).

For each block, the shares of the first pixel are securely protected. The remaining pixels have at least one valid share, as shown in Section~\ref{sec: the direct space-vacating method}. The original image can then be reconstructed with the process described in Section~\ref{sec:the correlation}.

\subsubsection{Data Extraction}
The receiver with the corresponding data hiding key \(K_D\) can extract the embedded data from the marked encrypted images and decrypt it. First, the receiver extracts \(S\), \(r\), \(n\), and \(ID\) from the first pixel of the first four blocks. Then, using Algorithm~\ref{alg:embedding space}, the available embedding space is determined within the marked encrypted image. The secret message is behind the 32 bits \(\bm{\overline{EP}}(ID, 0)\), \(\bm{\overline{EP}}(ID, 1)\), \(\bm{\overline{EP}}(ID, 2)\), and \(\bm{\overline{EP}}(ID, 3)\) in the embedding space. The receiver then extracts the embedded information and decrypts it using \(K_D\).

\subsection{An Example of the Proposed High-Capacity Scheme}
\begin{figure*}[!t]
    \centering
        \includegraphics[width=0.9\linewidth]{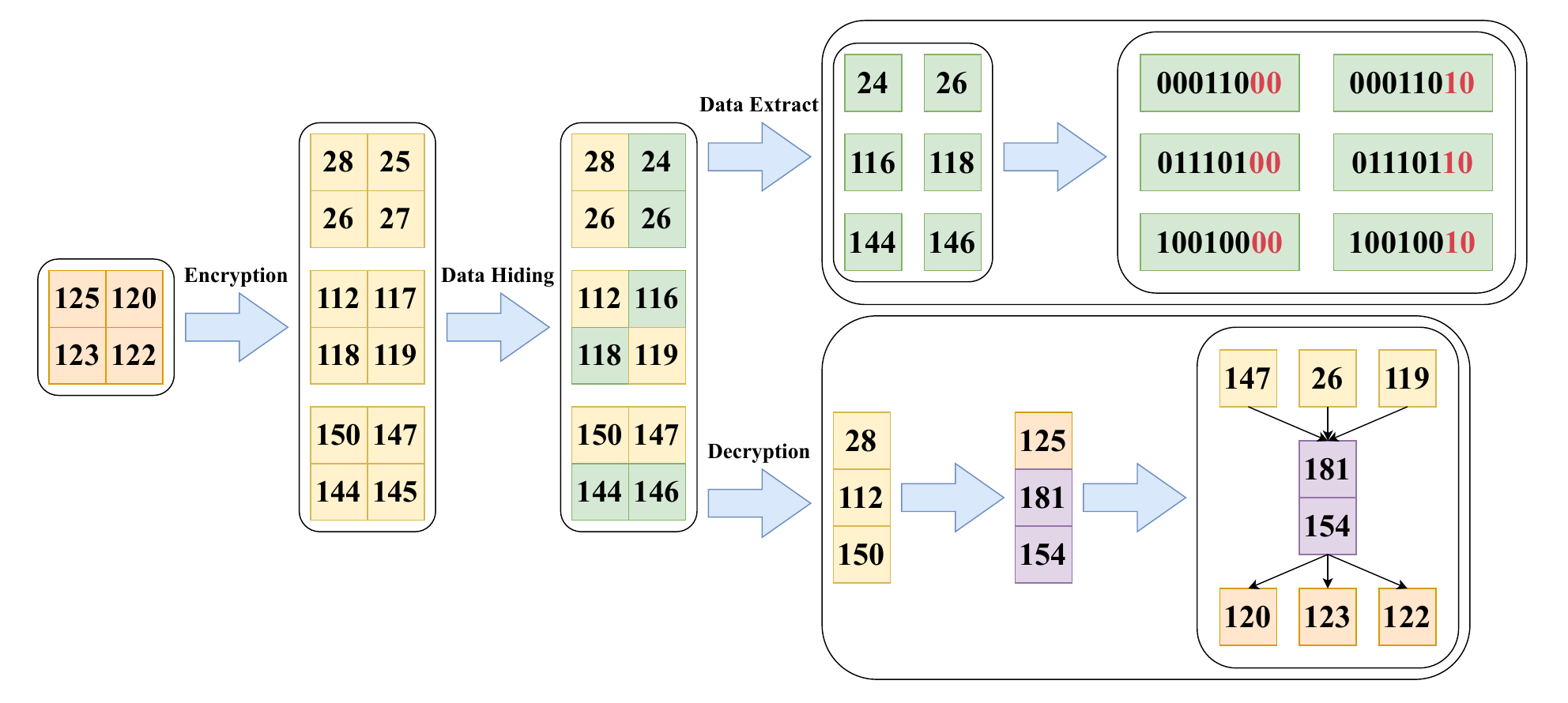}  
    \caption{An example of the proposed high-capacity scheme.}
    \label{Fig:example1}
\end{figure*}

To illustrate the proposed scheme, we provide an example in Fig.~\ref{Fig:example1}. The content owner sets \((r, n) = (3, 3)\) and \(S = 2\). Using the key \(K_E\), three distinct nonzero integers \(x(0), x(1), x(2)\) and two random numbers \(a(0), a(1)\) are generated. With these parameters, the content owner performs \((3, 3)\)-threshold block-based secret sharing on the original image, with block size \(2 \times 2\).

The encrypted blocks are computed as follows. First, the content owner derives the equation based on Eq.~(\ref{eq:used Shamir}):
\begin{align*}
    F_0(j, k, x) &= \bm{B}_0(j, k) \oplus 181 x \oplus 154 x^2 \mod p(y),
\end{align*}
For the \(0\)-th encrypted block, the content owner sets \(x = 21\):
\begin{align*}
    F_0(j, k, 21) &= \bm{B}_0(j, k) \oplus 181 \times 21 \oplus 154 \times 21^2 \mod p(y),
\end{align*}
The pixel values in the \(0\)-th encrypted block are calculated as follows:
\begin{align*}
    F_0(0, 0, 21) &= \bm{B}_0(0, 0) \oplus 181 \times 21 \oplus 154 \times 21^2 \mod p(y)\\
    &= 125 \oplus 181 \times 21 \oplus 154 \times 21^2 \mod p(y)\\
    &= 28,\\
    F_0(0, 1, 21) &= \bm{B}_0(0, 1) \oplus 181 \times 21 \oplus 154 \times 21^2 \mod p(y)\\
    &= 120 \oplus 181 \times 21 \oplus 154 \times 21^2 \mod p(y)\\
    &= 25,\\
    F_0(1, 0, 21) &= \bm{B}_0(1, 0) \oplus 181 \times 21 \oplus 154 \times 21^2 \mod p(y)\\
    &= 123 \oplus 181 \times 21 \oplus 154 \times 21^2 \mod p(y)\\
    &= 26,\\
    F_0(1, 1, 21) &= \bm{B}_0(1, 1) \oplus 181 \times 21 \oplus 154 \times 21^2 \mod p(y)\\
    &= 122 \oplus 181 \times 21 \oplus 154 \times 21^2 \mod p(y)\\
    &= 27.
\end{align*}
Similarly, the first and second encrypted blocks can be computed.

For the data hider, the following values are set for the embedding space:
\begin{align*}
    \bm{EP}(0) &= \{25, 26, 27\},\\
    \bm{EP}(1) &= \{117, 118, 119\},\\
    \bm{EP}(2) &= \{147, 144, 145\}.
\end{align*}
Using Algorithm~\ref{alg:embedding space}, the embedding space is determined as follows:
\begin{align*}
    \overline{\bm{EP}}(0) &= \{25, 27\},\\
    \overline{\bm{EP}}(1) &= \{117, 118\},\\
    \overline{\bm{EP}}(2) &= \{144, 145\}.
\end{align*}
As shown in Fig.~\ref{Fig:example1}, the green pixels represent the embedding space, while the yellow pixels are reserved for the original pixel values. The secret message, encrypted as "0001", is embedded into the embedding space from the least significant bit plane to the most significant bit plane for each encrypted block. Since the encrypted pixel values in the embedding space are meaningless, the data hider uses bit replacement to directly embed the secret message. For the \(0\)-th encrypted block, \(\bm{EB}_0(0, 1), \bm{EB}_0(1, 1)\) are used for data hiding. "00" is first embedded into the least significant bit plane, followed by "01" into the second least significant bit plane.

For the receiver, decryption or data extraction can be done based on the keys they possess.
A receiver with $K_E$ can decrypt the image block-by-block. Together with the pixel shares $\{ 28,112,150 \}$, the receiver uses Lagrangian interpolation to obtain the original value $125$ of the first pixel and two random numbers \(\{181,154\}\) of the block. The remaining pixel values in the block can be calculated according to Eq.(\ref{eq:decrypt Shamir}) as follows:

\begin{align*}
    \bm{B}_{0}(0,1) &= 147 \oplus  (181\times 21 \oplus 154 \times 21^2) \mod p(y)\\ &= 120,\\
    \bm{B}_{0}(0,2) &= 26 \oplus  (181\times 21 \oplus 154 \times 21^2) \mod p(y)\\ &= 123,\\
    \bm{B}_{0}(0,3) &= 119 \oplus  (181\times 21 \oplus 154 \times 21^2) \mod p(y)\\ &= 122.\\
\end{align*}

A receiver with \(K_D\) can extract the secret message from a specified location. The receiver first obtains \(\bm{EP}(0)\) and then uses Algorithm~\ref{alg:embedding space} to determine \(\overline{\bm{EP}}(0)=\{24,26\}\). The secret message is embedded in the lower bit planes of \(\overline{\bm{EP}}(0)\). The receiver extracts "00" from the least significant bit plane and extracts "01" from the second least significant bit plane. Finally, the receiver uses \(K_D\) to decrypt "0001" and obtain the secret message.

\section{The Proposed Size-Reduced RDH-EI Scheme}\label{sec:scheme II}
Based on Section~\ref{sec: image-shrinking-based space-vacating method}, an RDH-EI scheme whose encrypted images are size-reduced can be proposed. We encrypt the original image block-by-block and then discard part pixels of each encrypted image to form size-reduced ones. This can effectively reduce the occupancy of transmission bandwidth.

\subsection{Content Owner}
The content owner encrypts the original image using an encryption key $K_E$ to generate $n$ encrypted images. The encryption method is described in Section~\ref{sec: encryption}. The encrypted images are still the same size as the original image in this phase. After obtaining the encrypted images, we discard some pixels and resize them to form the size-reduced encrypted images. Denote that $\bm{WB}$ is an array utilized to store some encrypted blocks of an encrypted image, and $\bm{FP}$ is an array utilized to store the first pixels in some blocks of an encrypted image. The process is described as follows. 
\begin{enumerate}
    \item For $\bm{EB}_i$, if $ID$ of the corresponding encrypted image is in $\{i \mod n, i + 1 \mod n, \cdots, i + r - 2 \mod n\}$, store its first pixel in $\bm{FP}$. If not, store the whole block in $\bm{WB}$.
    \item Repeat step 1) for all blocks of $ID$-th encrypted image.
    % 计算M' 和N'
    \item Calculate the total pixel numbers $TP$ as $|\bm{FP}|+BS \times |\bm{WB}|$, where $|x|$ denotes the size of $x$. Then calculate $M'=\lceil \sqrt{\frac{TP}{BS}}\rceil \times S$ as  and set $N' = M'$.
    % 填充小尺寸图像
    \item Place the blocks of $\bm{WB}$ into $\bm{EI}'$ in raster order. For the pixels of $\bm{FP}$, first fill blocks in raster order, then insert these blocks into $\bm{EI}'$ in raster order. Finally, the remaining pixels in $\bm{EI}'$ are assigned random values.
    \item Embed parameters $S$, $r$, $n$, and $ID$ into the last four random number pixels of $\bm{EI}'$.
    \item Repeat step 1)-5) for all encrypted images.
\end{enumerate}

After this process, the final $n$ size-reduced encrypted images $\bm{EI}'$ are generated and sent to $n$ different data hiders.

\subsection{Data Hider}
When the $\beta$-th data hider receives a size-reduced encrypted image $\bm{EI}'({\beta})$, he/she can independently embed the secret message into the image to generate the ${\beta}$-th marked encrypted image $\bm{MI}({\beta})$. Firstly, the data hider extracts the last four pixels of the encrypted image to obtain the parameters $S$, $r$, $n$, and $ID$. The data hider calculates $|\bm{WB}|$ for each size-reduced encrypted image using algorithm~\ref{alg:wb number}. Subsequently, the encrypted image $\bm{EI}'({\beta})$ is partitioned into non-overlapping blocks of size $S\times S$. For the former $|\bm{WB}|$ blocks, it calculates the prediction errors and compresses them using arithmetic coding within each block as described in Section~\ref{sec: image-shrinking-based space-vacating method}. Finally, the vacated space together with the random numbers is utilized for data embedding.

\begin{algorithm}[!t]
\caption{\emph{Count the size of $\bm{WB}$ by ID}}
\label{alg:wb number}
\begin{algorithmic}[1]
\renewcommand{\algorithmicrequire}{\textbf{Initialize:}} 
\renewcommand{\algorithmicensure}{\textbf{Output:}}

\Require The number of image blocks $BN$, secret sharing threshold $r$, the number of encrypted images $n$, encrypted image identity $ID$.
\Ensure : The size $C$ of $\bm{WB}$

\State Initialize $SP = ID-r+2$.
\State Initialize $Cycle = \lfloor (BN-SP)/n \rfloor$.
\State Initialize $C = Cycle\times (n-r+1)$.
\If{$ID > r-2$ :}
    \State $C = C+ID-(r-2)$.
\EndIf
\If{$(BN-SP) \mod n > r-1$:}
    \State $C = C+(BN-SP) \mod n - (r-1)$.
\EndIf

\end{algorithmic}
\end{algorithm}

\begin{algorithm}[!t]
\caption{\emph{Count the size of $\bm{FP}$ by ID}}
\label{alg:fp number}
\begin{algorithmic}[1]
\renewcommand{\algorithmicrequire}{\textbf{Initialize:}} 
\renewcommand{\algorithmicensure}{\textbf{Output:}}

\Require The number of image blocks $BN$, secret sharing threshold $r$, the number of encrypted images $n$, encrypted image identity $ID$.
\Ensure : The size $D$ of $\bm{FP}$

\State Initialize $SP = ID-r+2$.
\State Initialize $Cycle = \lfloor (BN-SP)/n \rfloor$.
\State Initialize $D = Cycle\times (r-1)$.
\If{$ID <= r-2$ :}
    \State $D =D+ ID-(r-2)$.
\EndIf
\If{$(BN-SP) \mod n <= r-1$:}
    \State $D =D+ (BN-SP) \mod n$.
\EndIf

\end{algorithmic}
\end{algorithm}

The embedding space in this scheme refers to the pixels within the former $|\bm{WB}|$ image blocks, excluding the first pixel in each block, together with the pixels set as random numbers.

\subsection{Receiver}
A receiver equipped with the encryption key $ K_E $ and $ r $ marked encrypted images can reconstruct the original image. In contrast, a receiver holding the data hiding key $ K_D $ can recover the embedded data from the marked encrypted images they have received.

\subsubsection{Decryption}
After acquiring $r$ marked encrypted images $\bm{MI}(\gamma_0)$, $\bm{MI}(\gamma_1), \cdots, \bm{MI}(\gamma_{r-1})$, the receiver first restores the size-reduced encrypted images and then reconstruct $\bm{I}$.

To initiate the restoration of the encrypted images, the receiver first extracts parameters from the last four pixels of each image. Specifically, the block size $S$, the threshold $r$
the number of generated shares $n$, and the identity of the marked encrypted image $ID$. Subsequently, the receiver calculates $|WB|$ and proceeds to extract the side information bitstream $SI$ from the former $|WB|$ image blocks, excluding the first pixel in each block. 
According to $SI$, the length of $CB$ and the numbers of each error value can be obtained based on binary. At the position after $SI$, the receiver extracts $|CB|$ bits and converts them into error values using arithmetic decoding.

\noindent \textbf{Size-Reduced Encrypted Image Recovery.}
After using arithmetic decoding to recover error blocks of the size-reduced encrypted image, the first pixel remains unchanged for blocks. Leveraging the first pixels and the prediction errors, the encrypted image can be reconstructed using MED.

For each error block $\bm{ERB}_i$, its first pixel value $\bm{ERB}_i(0,0)$ is still the same as value in $\bm{EB}'_i(0,0)$. The size-reduced encrypted image recovery process proceeds as follows, where $p(j,k) = \bm{EB}'_i(j,k)$ and $e(j,k)=\bm{ERB}_i(j,k)$:
\begin{enumerate}
    \item Iterate over $j$ from 0 to $S-1$ and $k$ from 0 to $S-1$, excluding the case where $j=k=0$, use MED to compute $\hat{p}(j,k)$ and subsequently obtain $ p(j,k) $ as $p(j,k) = \hat{p}(j,k) + e(j,k)$ within each iteration. Then, ${\bm{EB}'}_i$ is obtained. 
    \item Repeat step 1) for the former $|\bm{WB}|$ blocks to achieve the recovery of the size-reduced encrypted image.
\end{enumerate}

Then, for each size-reduced encrypted image, the former $|\bm{WB}|$ blocks are extracted to form $\bm{WB}$. Using algorithm~\ref{alg:fp number} to calculate the size of $\bm{FP}$, the next $|\bm{FP}|$ pixels are extracted to form $\bm{FP}$. To recover the original image, $\bm{WB}$s and $\bm{FP}s$ are utilized as follows for decryption.
\begin{enumerate}
    \item For the $i$-th original block $\bm{B}_i$, traversal $ID$ from $\gamma_0$ to $\gamma_{r-1}$, if $ID$ is in $\{i \mod n, i + 1 \mod n, \cdots, i + r - 2 \mod n\}$, extract a pixel from $\bm{FP}$ of the $ID$-th size-reduced encrypted image in order. If not, extract a block from $\bm{WB}$ of the $ID$-th size-reduced encrypted image in order. Denote the pixels as $\mathcal{C}$ and the blocks as $\mathcal{D}$.
    \item For the first pixel of $\bm{B}_i$, $\mathcal{C}$ and the first pixels of $\mathcal{D}$ are utilized for decryption.
    \item For any of the remaining pixels in $\bm{B}_i$, the pixel at the corresponding position in any block of $\mathcal{D}$ are utilized for decryption.
    \item Repeat 1)-3) for $i$ from $0$ to $BN-1$.
\end{enumerate}
Finally, the original image $\bm{I}$ is recovered.

\subsubsection{Data Extraction}
The receiver, using the data hiding key $ K_D $, can retrieve the embedded data from the received marked encrypted image and decrypt it. First, the block size $ S $ is extracted from the fourth-to-last pixel in the last row. Then, the position of the first available space in the marked encrypted image is identified using $ SI $. Once this position is located, the receiver can extract all the embedded data. Finally, the message is obtained by decrypting the extracted data with $ K_D $.

\subsection{An Example of the Proposed Size-Reduced Scheme}
\begin{figure*}[!t]
    \centering
        \includegraphics[width=1\linewidth]{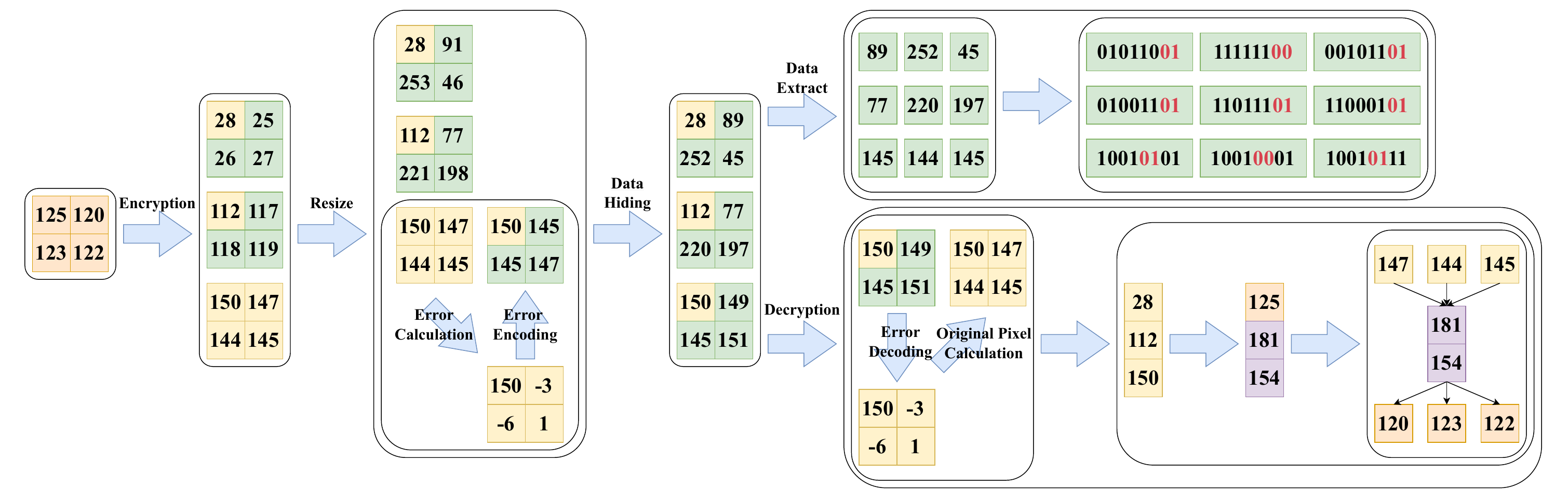}  
    \caption{An example of the size-reduced scheme.}
    \label{Fig:example2}
\end{figure*}

In this part, an example is performed in Fig.~\ref{Fig:example2} to illustrate the differences between this RDH-EI scheme and the previous one.

After encryption, the content owner discards a few pixels of each encrypted image according to Section~\ref{sec: image-shrinking-based space-vacating method}. For the $0$-th and $1$-th encrypted blocks, the content owner only saves its first pixels $\{ 28, 112\}$. And for $\bm{EB}_0$ in $2$-th encrypted image, it saves the whole block $\{150,147,144,145\}$.

After discarding certain pixels, the content owner performs a resize operation due to the scattered distribution of the remaining pixels on the encrypted image after discarding some pixels. For each encrypted block, the content owner calculates its $\bm{FP}$ and $\bm{WB}$. Then $M'$ and $N'$ are calculated. To form a size-reduced encrypted image with a size of $M'\times N'$, the blocks in $\bm{WB}$ are put to the former, and the pixels in $\bm{FP}$ are put to fill a block in raster order. The remaining pixel is set as a random number.
In real scenarios, the generated images are smaller and the parameters are embedded into the random pixel values at the end of the size-reduced encrypted images. We ignore this operation in the example for simplicity.

To create embedding space, the data hider applies the space-vacating operation. The data hider first calculates the prediction errors of blocks in $\bm{WB}$ using MED. Then the errors are compressed using arithmetic coding. Due to the limited number of error values in the example provided, arithmetic coding cannot compress them effectively. Therefore, we arbitrarily use bit "111001" as compression bits for the errors to simulate the effect. Additionally, because the image is too small, it cannot store $SI$, so we will ignore the storage of $SI$ here. MED is utilized for this block $\{150,147,144,145\}$:
\begin{align*}
    \hat{p}(0,1) &= p(0,0) =  150,\\
    \hat{p}(1,0) &= p(0,0) = 150,\\
    \hat{p}(1,1) &= p(1,0) = 144.
\end{align*}
Prediction errors can be calculated as:
\begin{align*}
    e(0,1)&=p(0,1)-\hat{p}(0,1) = -3,\\
    e(1,0)&=p(1,0)-\hat{p}(1,0) = -6,\\
    e(1,1)&=p(1,1)-\hat{p}(1,1) = 1.
\end{align*}

The errors $\{-3, -6, 1\}$ are compressed into $CB = "111001"$, and the length of $CB$ is recorded in $SI$ as "000110". For each error value in the range $[-255, 255]$, the data hider uses 6 bits to record its occurrence. Since the only error values present are -3, -6, and 1, "000000" is appended to $SI$ for the other error values, while "000001" is added for -3, -6, and 1.
The embedding space is the green pixel as shown in Fig.~\ref{Fig:example2}. The data hider puts the $CB$ first and then embeds the secret message.

During the decryption process, the receiver restores each block. 
First, the receiver calculates $|WB|$ and extracts $CB$ from the embedding space of each block. Using this bitstream, the receiver recovers the prediction errors $\{-3,-6,1\}$. Using MED, the $2$-th block can be recovered. Then, $\bm{FP}$s and $\bm{WB}$s can be calculated. $\bm{FP}(0) = \{28\}$, $\bm{FP}(1) = \{112\}$ and $\bm{WB}(2)=\{150,147,144,145\}$. $28$, $112$ and $150$ are utilized to decrypt $125$ along with two random numbers $181$ and $154$ using Lagrangian interpolation. The remaining pixel shares $\{147,144,145\}$ are utilized to decrypt the remaining pixels according to the novel decryption.

\section{Experimental Results} \label{sec: Experimental Results}

\begin{figure*}[!t]
    \centering
    \begin{subfigure}{0.15\textwidth}
        \includegraphics[width=\linewidth]{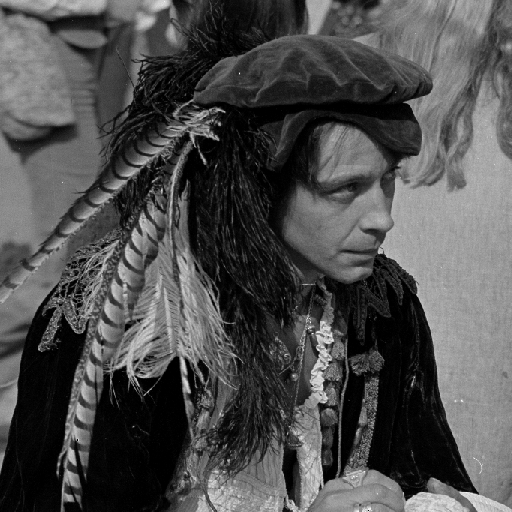}
        \caption{}
    \end{subfigure}
    \hfill
    \begin{subfigure}{0.15\textwidth}
        \includegraphics[width=\linewidth]{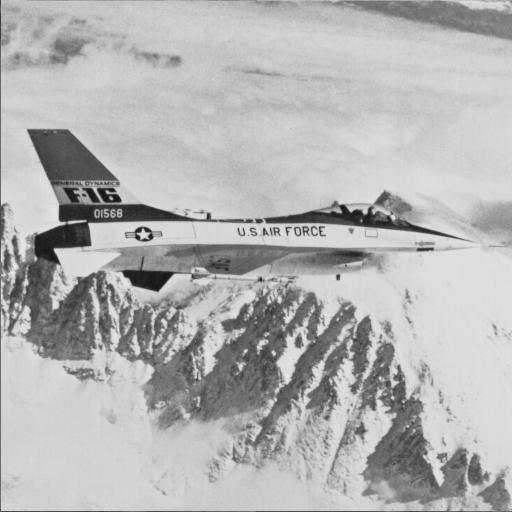}
        \caption{}
    \end{subfigure}
    \hfill
    \begin{subfigure}{0.15\textwidth}
        \includegraphics[width=\linewidth]{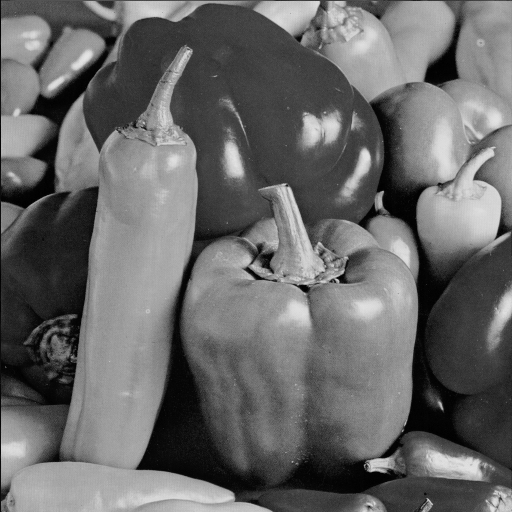}
        \caption{}
    \end{subfigure} 
    \hfill
    \begin{subfigure}{0.15\textwidth}
        \includegraphics[width=\linewidth]{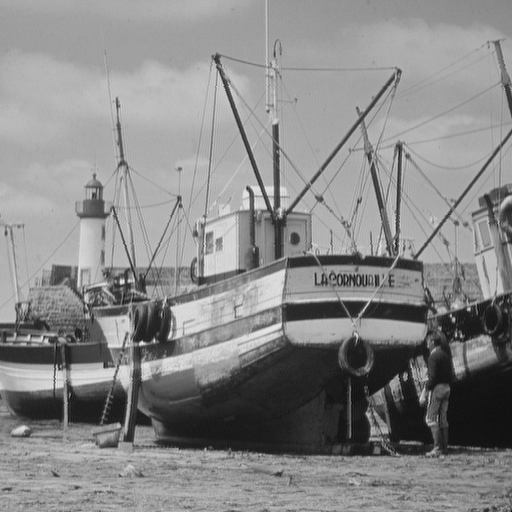}
        \caption{}
    \end{subfigure} 
    \hfill
    \begin{subfigure}{0.15\textwidth}
        \includegraphics[width=\linewidth]{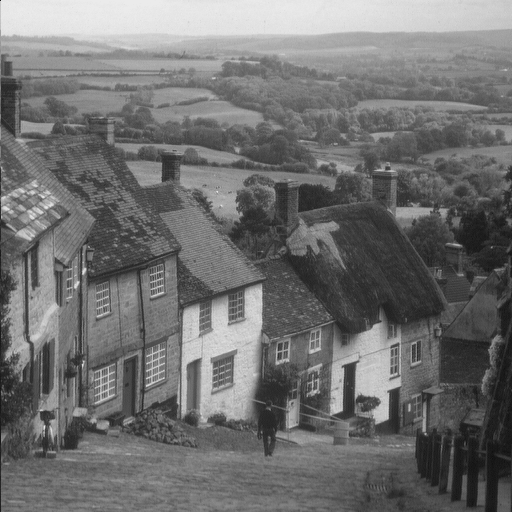}
        \caption{}
    \end{subfigure}
    \hfill
    \begin{subfigure}{0.15\textwidth}
        \includegraphics[width=\linewidth]{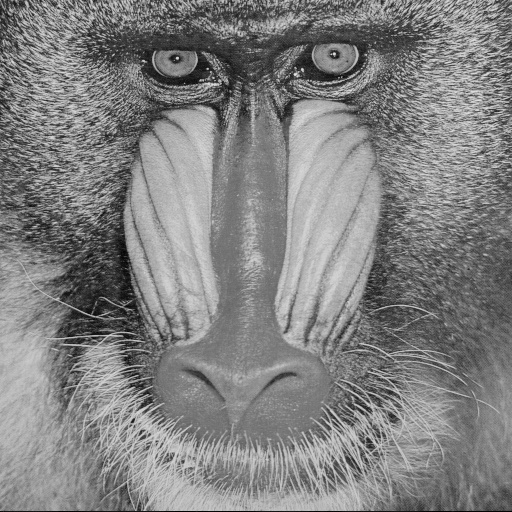}
        \caption{}
    \end{subfigure} 
    \caption{Six test images: (a) \textit{Man}; (b) \textit{Jetplane}; (c) \textit{Peppers}; (d) \textit{Boat}; (e) \textit{Goldhill}; (f) \textit{Baboon}.}
    \label{Fig:test images}
\end{figure*}

This section evaluates two proposed schemes: the high-capacity RDH-EI, which directly generates embedding space, and the size-reduced RDH-EI, which produces size-reduced encrypted images. The evaluation uses six typical grayscale images shown in Fig.~\ref{Fig:test images} and two datasets, BOSSBase~\cite{bas_break_2011} and BOWS-2~\cite{bas2017image}, each containing 10,000 grayscale images of size $512 \times 512$. In the following tables and figures, "ours(capacity)" represents "the high-capacity scheme", and "ours(size)" represents "the size-reduced scheme".
\subsection{Visual Evaluation}
\begin{figure}[!t]
    \centering
    \begin{subfigure}{0.09\textwidth}
        \includegraphics[width=\linewidth]{picture/baboon.png}
        \caption{}
    \end{subfigure}
    %\hfill
    \begin{subfigure}{0.09\textwidth}
        \includegraphics[width=\linewidth]{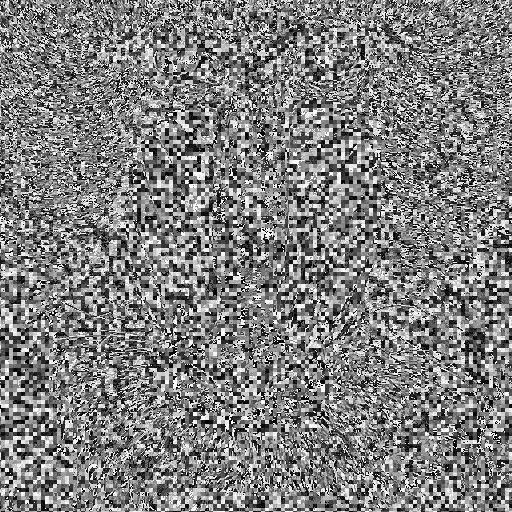}
        \caption{}
    \end{subfigure}
    %\hfill
    \begin{subfigure}{0.09\textwidth}
        \includegraphics[width=\linewidth]{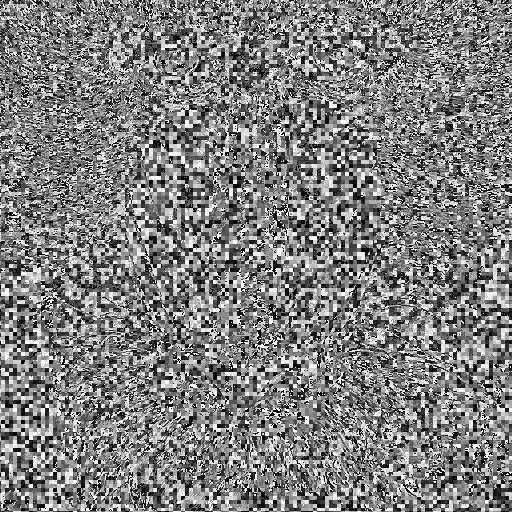}
        \caption{}
    \end{subfigure}
    %\hfill
    \begin{subfigure}{0.09\textwidth}
        \includegraphics[width=\linewidth]{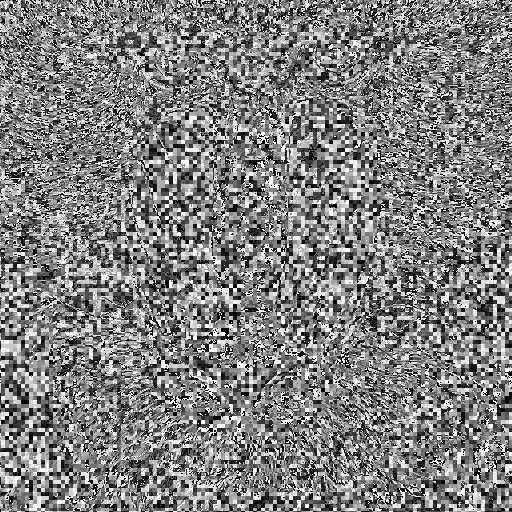}
        \caption{}
    \end{subfigure}
    %\hfill
    \begin{subfigure}{0.09\textwidth}
        \includegraphics[width=\linewidth]{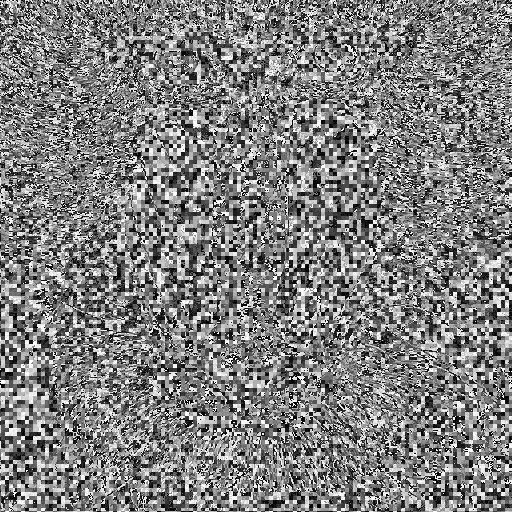}
        \caption{}
    \end{subfigure}
    \medskip
    \begin{subfigure}{0.09\textwidth}
        \includegraphics[width=\linewidth]{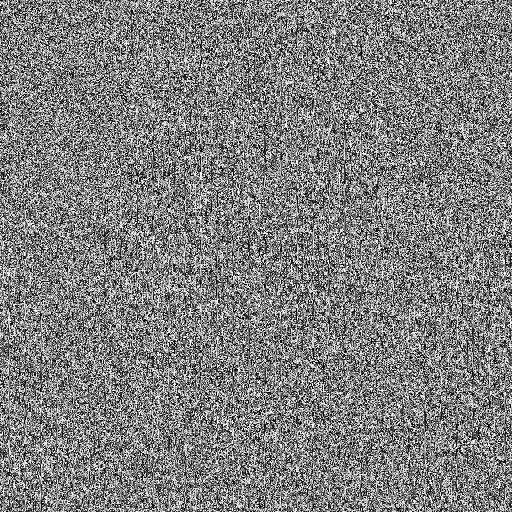}
        \caption{}
    \end{subfigure}
    %\hfill
    \begin{subfigure}{0.09\textwidth}
        \includegraphics[width=\linewidth]{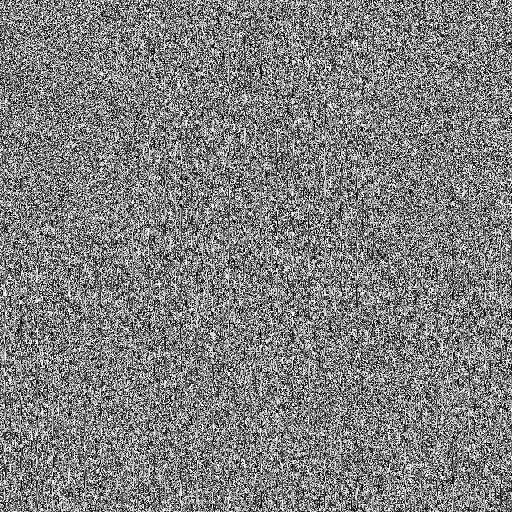}
        \caption{}
    \end{subfigure}
    %\hfill
    \begin{subfigure}{0.09\textwidth}
        \includegraphics[width=\linewidth]{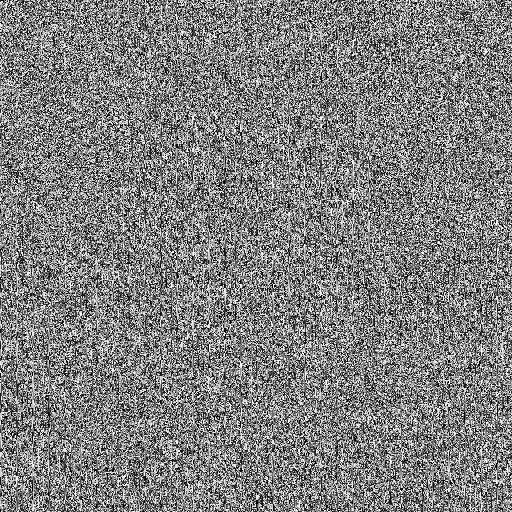}
        \caption{}
    \end{subfigure}
    %\hfill
    \begin{subfigure}{0.09\textwidth}
        \includegraphics[width=\linewidth]{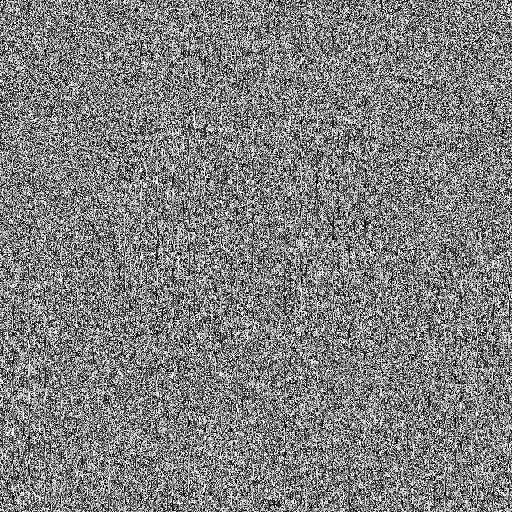}
        \caption{}
    \end{subfigure}
    %\hfill
    \begin{subfigure}{0.09\textwidth}
        \includegraphics[width=\linewidth]{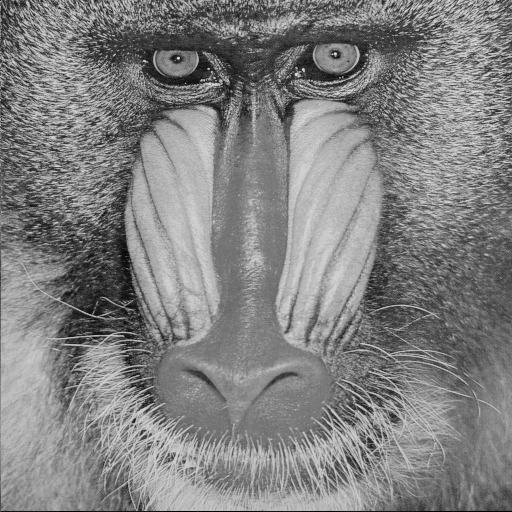}
        \caption{}
    \end{subfigure}

    \begin{subfigure}{0.09\textwidth}
        \includegraphics[width=\linewidth]{picture/jetplane.png}
        \caption{}
    \end{subfigure}
    %\hfill
    \begin{subfigure}{0.09\textwidth}
        \includegraphics[width=\linewidth]{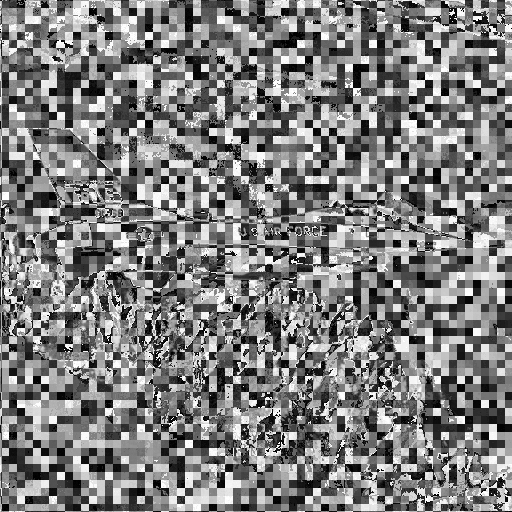}
        \caption{}
    \end{subfigure}
    %\hfill
    \begin{subfigure}{0.09\textwidth}
        \includegraphics[width=\linewidth]{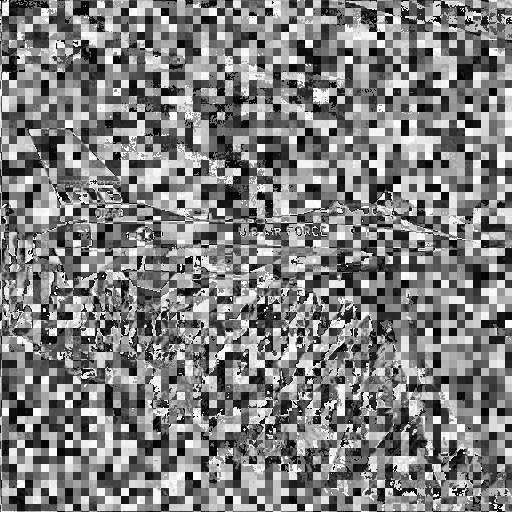}
        \caption{}
    \end{subfigure}
    %\hfill
    \begin{subfigure}{0.09\textwidth}
        \includegraphics[width=\linewidth]{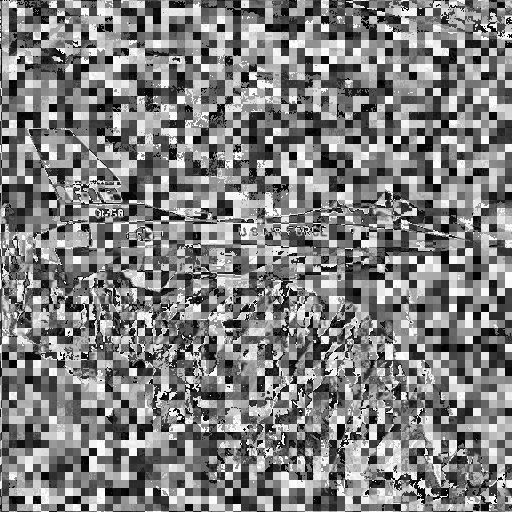}
        \caption{}
    \end{subfigure}
    %\hfill
    \begin{subfigure}{0.09\textwidth}
        \includegraphics[width=\linewidth]{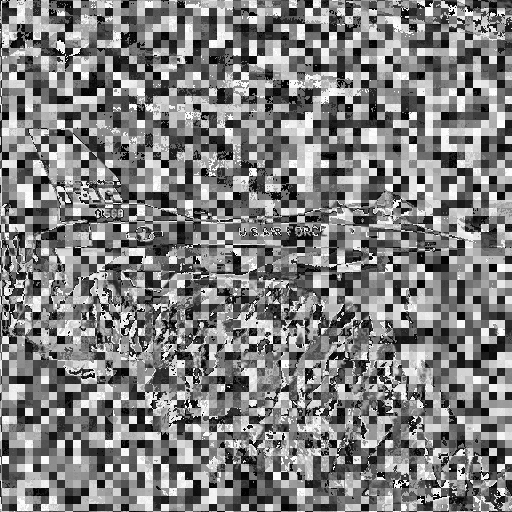}
        \caption{}
    \end{subfigure}
    \medskip
    \begin{subfigure}{0.09\textwidth}
        \includegraphics[width=\linewidth]{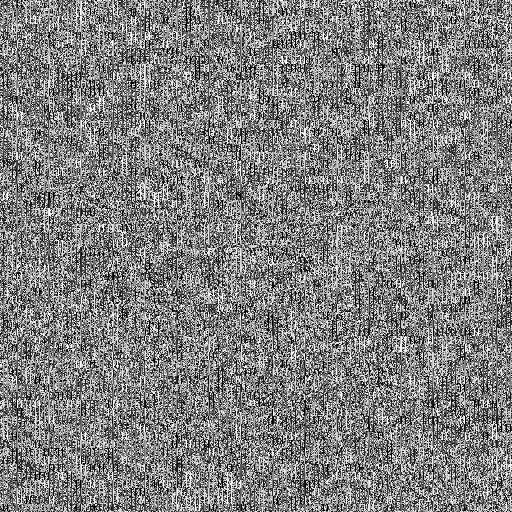}
        \caption{}
    \end{subfigure}
    %\hfill
    \begin{subfigure}{0.09\textwidth}
        \includegraphics[width=\linewidth]{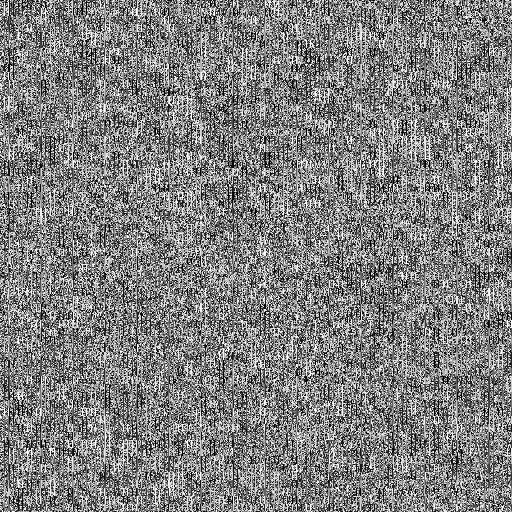}
        \caption{}
    \end{subfigure}
    %\hfill
    \begin{subfigure}{0.09\textwidth}
        \includegraphics[width=\linewidth]{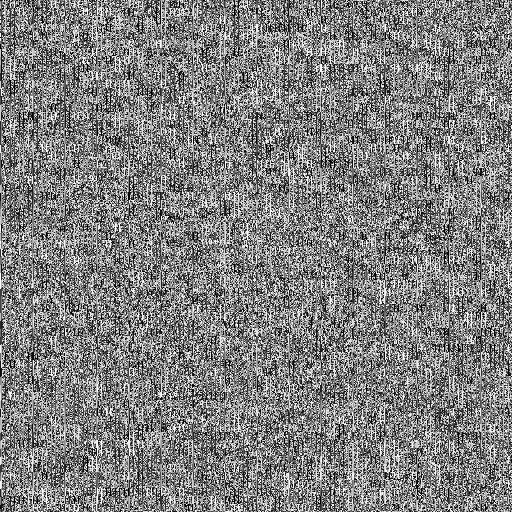}
        \caption{}
    \end{subfigure}
    %\hfill
    \begin{subfigure}{0.09\textwidth}
        \includegraphics[width=\linewidth]{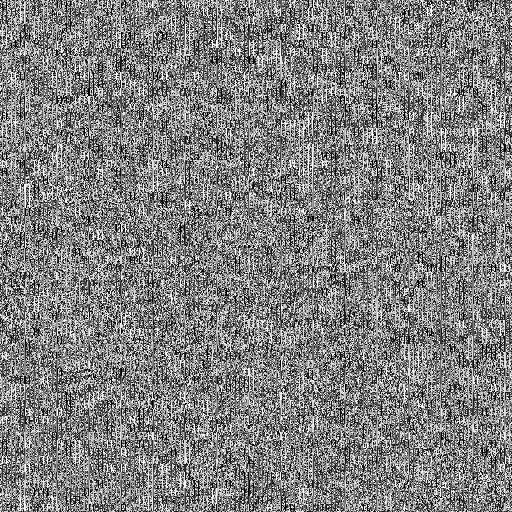}
        \caption{}
    \end{subfigure}
    %\hfill
    \begin{subfigure}{0.09\textwidth}
        \includegraphics[width=\linewidth]{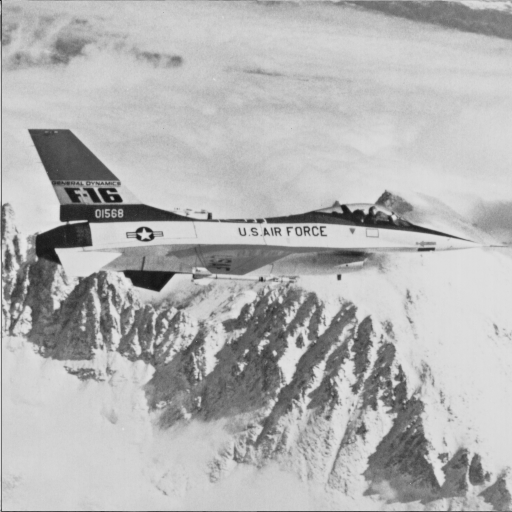}
        \caption{}
    \end{subfigure}
    \caption{Simulation experiments of $(4,4)$-threshold. (a) Image \textit{Baboon}; (b)-(e) four encrypted images with size $4 \times 4$; (f)-(i) four marked encrypted images; (j) recovered image with $PSNR \rightarrow +\infty$; (k) image \textit{Jetplane}; (l)-(o) four encrypted images with size $8 \times 8$; (p)-(s) Four marked encrypted images; (t) recovered image with $PSNR \rightarrow +\infty$.}
    \label{Fig:simulation result1}
    \vspace{-5pt}
\end{figure}
In both schemes, there are three key parameters: $S$, $n$, and $r$. The parameter $S$ represents the block size, which significantly influences both the visual impact of the encrypted images and the data embedding performance. Additionally, $n$ denotes the total number of encrypted images, and $r$ indicates the minimum number of encrypted images required to successfully recover the original image. Notably, $n$ and $r$ have minimal impact on the visual performance of the encrypted images generated by the high-capacity scheme but do affect the size-reduced scheme.

In the experiments of the high-capacity scheme, we fix $r$ and $n$ as 4, while considering $S$ to be either 4 or 8.
Fig.~\ref{Fig:simulation result1} shows the simulation results for the image \textit{Baboon} with a size of $4 \times 4$ and the image \textit{Jetplane}  with a size of $8 \times 8$. By comparing Figs.~\ref{Fig:simulation result1} (b)-(e) and (l)-(o), we observe that larger block sizes lead to larger artifacts. Furthermore, after embedding data into the encrypted images, the marked encrypted images exhibit no visual artifacts, resulting in superior visual effects.

\begin{figure}[!t]
    \centering
    \begin{subfigure}{0.09\textwidth}
        \includegraphics[width=\linewidth]{picture/man.png}
        \caption{}
    \end{subfigure}
    \begin{subfigure}{0.066\textwidth}
        \includegraphics[width=\linewidth]{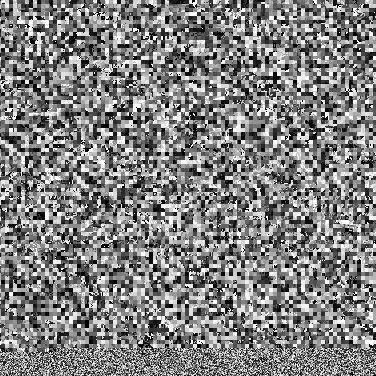}
        \caption{}
    \end{subfigure}
    \begin{subfigure}{0.066\textwidth}
        \includegraphics[width=\linewidth]{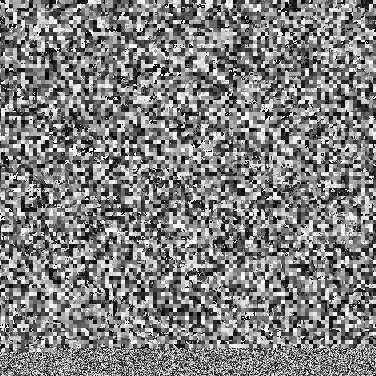}
        \caption{}
    \end{subfigure}
    \begin{subfigure}{0.066\textwidth}
        \includegraphics[width=\linewidth]{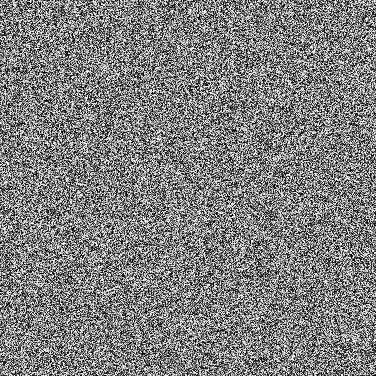}
        \caption{}
    \end{subfigure}
    \begin{subfigure}{0.066\textwidth}
        \includegraphics[width=\linewidth]{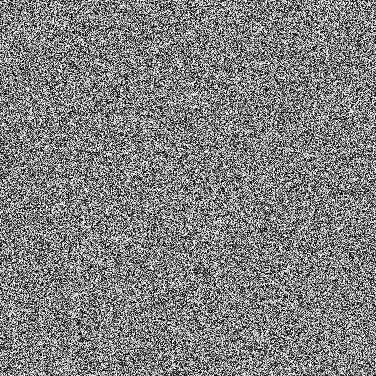}
        \caption{}
    \end{subfigure}
    \begin{subfigure}{0.09\textwidth}
        \includegraphics[width=\linewidth]{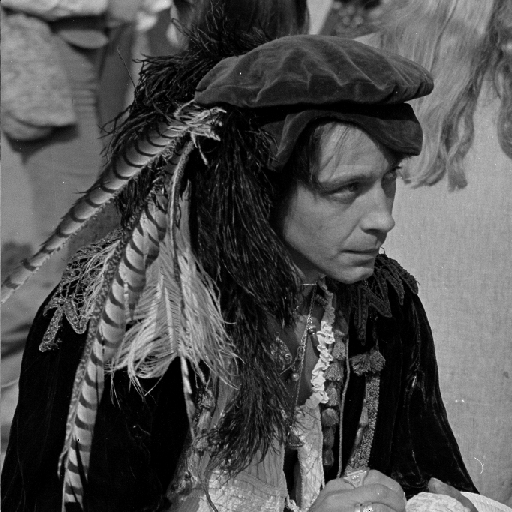}
        \caption{}
    \end{subfigure}

    \begin{subfigure}{0.09\textwidth}

        \includegraphics[width=\linewidth]{picture/peppers.png}
        \caption{}
    \end{subfigure}
    %\hfill
    \begin{subfigure}{0.065\textwidth}

        \includegraphics[width=\linewidth]{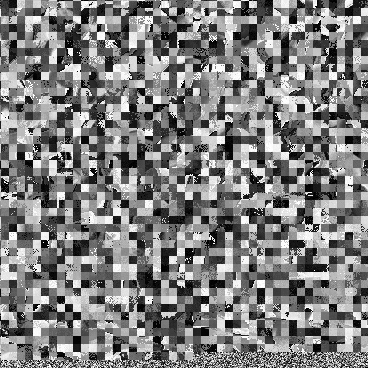}
        \caption{}
    \end{subfigure}
    %\hfill
    \begin{subfigure}{0.065\textwidth}

        \includegraphics[width=\linewidth]{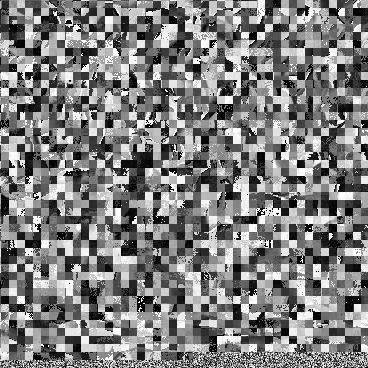}
        \caption{}
    \end{subfigure}
    \begin{subfigure}{0.065\textwidth}

        \includegraphics[width=\linewidth]{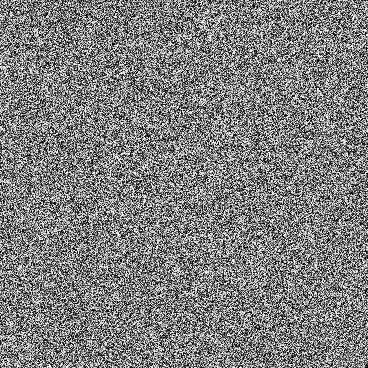}
        \caption{}
    \end{subfigure}
    %\hfill
    \begin{subfigure}{0.065\textwidth}

        \includegraphics[width=\linewidth]{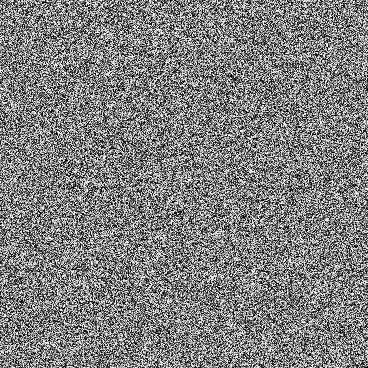}
        \caption{}
    \end{subfigure}
    \begin{subfigure}{0.09\textwidth}

        \includegraphics[width=\linewidth]{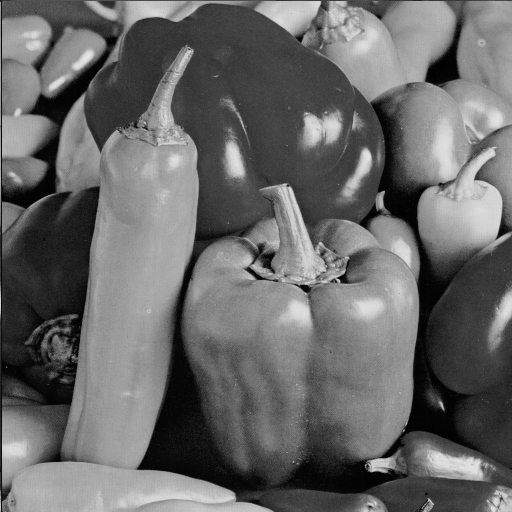}
        \caption{}
    \end{subfigure}
    \caption{Simulation experiments of $(2,2)$-threshold. (a) Image \textit{Man}; (b)-(c) two size-reduced encrypted images with size $4 \times 4$; (d)-(e) two size-reduced marked encrypted images; (f) recovered image with $PSNR \rightarrow +\infty$; (g) image \textit{Peppers}; (h)-(i) two size-reduced encrypted images with size $8 \times 8$; (j)-(k) two size-reduced marked encrypted images; (l) recovered image with $PSNR \rightarrow +\infty$.}
    \label{Fig:simulation result2}
    \vspace{-5pt}
\end{figure}

\begin{figure}[!t]
    \centering
    \begin{subfigure}{0.09\textwidth}
        \centering
        \includegraphics[width=\linewidth]{picture/boat.png}
        \caption{}
    \end{subfigure}
    %\hfill
    \begin{subfigure}{0.09\textwidth}
    \centering
        \includegraphics[width=0.55\linewidth]{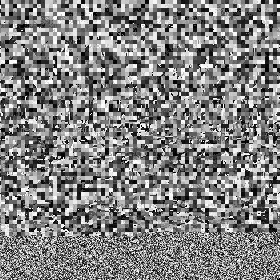}
        \caption{}
    \end{subfigure}
    %\hfill
    \begin{subfigure}{0.09\textwidth}
    \centering
        \includegraphics[width=0.55\linewidth]{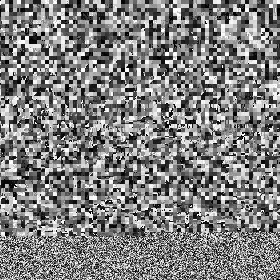}
        \caption{}
    \end{subfigure}
    %\hfill
    \begin{subfigure}{0.09\textwidth}
    \centering
        \includegraphics[width=0.55\linewidth]{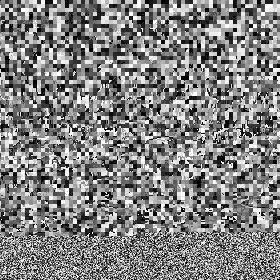}
        \caption{}
    \end{subfigure}
    %\hfill
    \begin{subfigure}{0.09\textwidth}
    \centering
        \includegraphics[width=0.55\linewidth]{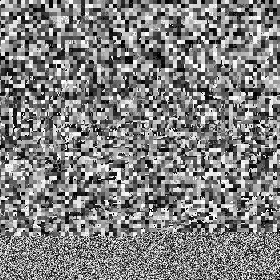}
        \caption{}
    \end{subfigure}
    \medskip
    \begin{subfigure}{0.09\textwidth}
    \centering
        \includegraphics[width=0.55\linewidth]{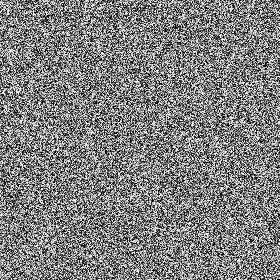}
        \caption{}
    \end{subfigure}
    %\hfill
    \begin{subfigure}{0.09\textwidth}
    \centering
        \includegraphics[width=0.55\linewidth]{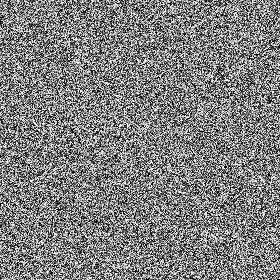}
        \caption{}
    \end{subfigure}
    %\hfill
    \begin{subfigure}{0.09\textwidth}
    \centering
        \includegraphics[width=0.55\linewidth]{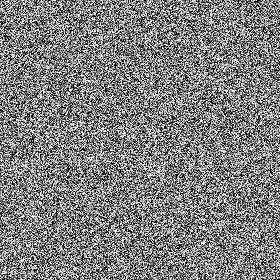}
        \caption{}
    \end{subfigure}
    %\hfill
    \begin{subfigure}{0.09\textwidth}
    \centering
        \includegraphics[width=0.55\linewidth]{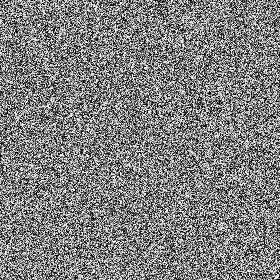}
        \caption{}
    \end{subfigure}
    %\hfill
    \begin{subfigure}{0.09\textwidth}
    \centering
        \includegraphics[width=\linewidth]{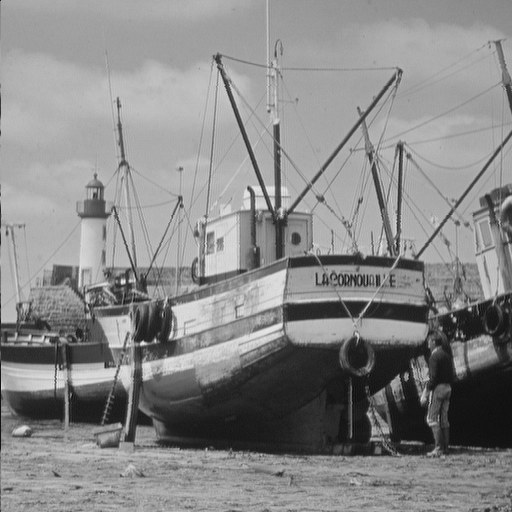}
        \caption{}
    \end{subfigure}

    \begin{subfigure}{0.09\textwidth}
    \centering
        \includegraphics[width=\linewidth]{picture/goldhill.png}
        \caption{}
    \end{subfigure}
    %\hfill
    \begin{subfigure}{0.09\textwidth}
    \centering
        \includegraphics[width=0.52\linewidth]{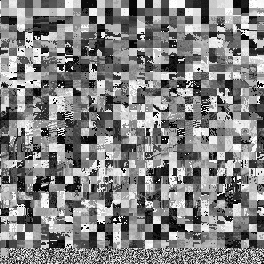}
        \caption{}
    \end{subfigure}
    %\hfill
    \begin{subfigure}{0.09\textwidth}
    \centering
        \includegraphics[width=0.52\linewidth]{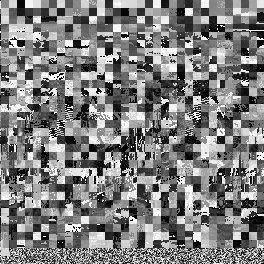}
        \caption{}
    \end{subfigure}
    %\hfill
    \begin{subfigure}{0.09\textwidth}
    \centering
        \includegraphics[width=0.52\linewidth]{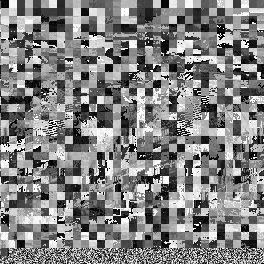}
        \caption{}
    \end{subfigure}
    %\hfill
    \begin{subfigure}{0.09\textwidth}
    \centering
        \includegraphics[width=0.52\linewidth]{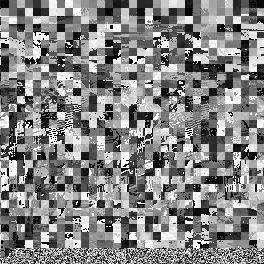}
        \caption{}
    \end{subfigure}
    \medskip
    \begin{subfigure}{0.09\textwidth}
    \centering
        \includegraphics[width=0.52\linewidth]{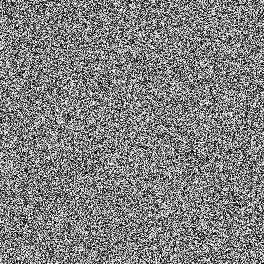}
        \caption{}
    \end{subfigure}
    %\hfill
    \begin{subfigure}{0.09\textwidth}
    \centering
        \includegraphics[width=0.52\linewidth]{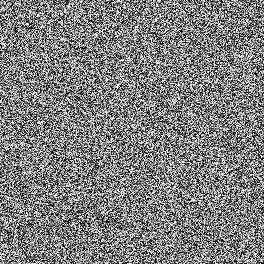}
        \caption{}
    \end{subfigure}
    %\hfill
    \begin{subfigure}{0.09\textwidth}
    \centering
        \includegraphics[width=0.52\linewidth]{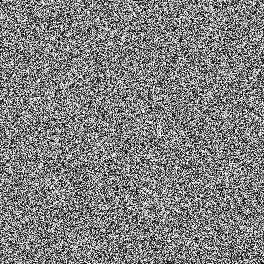}
        \caption{}
    \end{subfigure}
    %\hfill
    \begin{subfigure}{0.09\textwidth}
    \centering
        \includegraphics[width=0.52\linewidth]{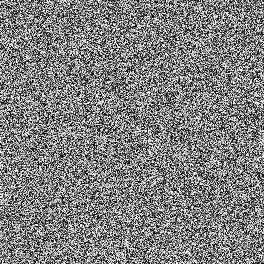}
        \caption{}
    \end{subfigure}
    %\hfill
    \begin{subfigure}{0.09\textwidth}
    \centering
        \includegraphics[width=\linewidth]{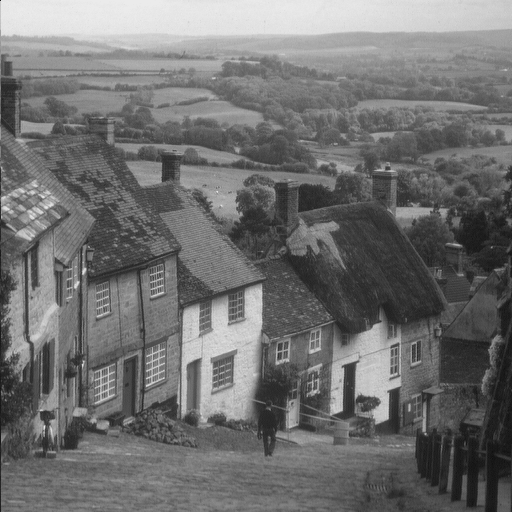}
        \caption{}
    \end{subfigure}
    \caption{Simulation experiments of $(4,4)$-threshold. (a) Image \textit{Boat}; (b)-(e) four size-reduced encrypted images with size $4 \times 4$; (f)-(i) four size-reduced marked encrypted images; (j) recovered image with $PSNR \rightarrow +\infty$; (k) image \textit{Goldhill}; (l)-(o) four size-reduced encrypted images with size $8 \times 8$; (p)-(s) four size-reduced marked encrypted images; (t) recovered image with $PSNR \rightarrow +\infty$.}
    \label{Fig:simulation result3}
    \vspace{-5pt}
\end{figure}

In the experiments for the size-reduced scheme, we consider $(r, n)$ values of either $(2, 2)$ or $(4, 4)$ and set $S$ to 4 or 8. Fig.~\ref{Fig:simulation result2} illustrates the simulation results for the images \textit{Man} under $(2,2)$ with a size of $4 \times 4$ and \textit{Peppers} with a size of $8 \times 8$. For the encrypted images of \textit{Man}, both $M'$ and $N'$ are 376, and for \textit{Peppers}, they are 368. The encrypted and marked encrypted images are smaller than the original images, with their sizes influenced by the parameters $r$, $n$, and $S$. The parameter $S$ has minimal impact on the size of the encrypted images when larger than or equal to 4.

Fig.~\ref{Fig:simulation result3} presents the simulation results under $(4,4)$ for the images \textit{Boat} with a size of $4 \times 4$ and \textit{Goldhill} with a size of $8 \times 8$. For the encrypted images of \textit{Boat}, both $M'$ and $N'$ are 280, while for \textit{Goldhill}, they are 264. 
Comparing encrypted images with the same block size but different $(r, n)$ values reveals that $r$ and $n$ significantly affect the size of encrypted images. When $r = n$, increasing $r$ reduces the size of the resulting encrypted images. 

\subsection{Data Embedding Capacity}

\begin{table}[h!]
\small
\caption{Embedding rates of the high-capacity scheme with different $r$, $n$ and $S$.}
\label{tab:ER1}
\setlength\tabcolsep{8pt}
\begin{tabular}{ccccccc}
\toprule
\textbf{$S$} & \textbf{$r$} & $n=2$ & $n=3$ & $n=4$ & $n=5$ & $n=6$ \\ 
\midrule
\multirow{5}{*}{4} & 2 & 3.75   & 2.5    & 1.875  & 1.5    & 1.25   \\
                   & 3 & --     & 5.0    & 3.75   & 3.0    & 2.5    \\
                   & 4 & --     & --     & 5.625  & 4.5    & 3.75   \\
                   & 5 & --     & --     & --     & 6.0    & 5.0    \\
                   & 6 & --     & --     & --     & --     & 6.25   \\
\multirow{5}{*}{8} & 2 & 3.9375 & 2.625  & 1.9688 & 1.575  & 1.3125 \\
                   & 3 & --     & 5.25   & 3.9375 & 3.15   & 2.625  \\
                   & 4 & --     & --     & 5.9063 & 4.725  & 3.9375 \\
                   & 5 & --     & --     & --     & 6.3    & 5.25   \\
                   & 6 & --     & --     & --     & --     & 6.5625 \\
\bottomrule
\end{tabular}
\end{table}

\begin{table}[!t]
    \small
    \begin{center}
        \caption{Embedding rates of the size-reduced scheme for different images with $(r,n)$ and $BS$.}
        \label{TAB:ER2}
        \setlength\tabcolsep{17pt}
        \begin{tabular}{cccc}
            \toprule
            Image & $(r,n) $& $S=4$ & $S=8$\\
            \midrule
            \multirow{2}{*}{\textit{Man}}   &$(2,2)$    & 1.5423    & \textbf{1.7091}  \\
                                            &$(4,4)$    & 1.2925    &  1.5910\\
            \multirow{2}{*}{\textit{Jetplane}}   &$(2,2)$    & 2.2953    & \textbf{2.5410}  \\
                                                &$(4,4)$    & 1.9692    &2.4081\\
            \multirow{2}{*}{\textit{Peppers}}    &$(2,2)$    & 1.7224    & \textbf{1.8924}  \\
                                                &$(4,4)$    & 1.4558    &1.7757\\
            \multirow{2}{*}{\textit{Boat}}       &$(2,2)$    & 1.8813    & \textbf{2.0970}  \\
                                                &$(4,4)$    & 1.5977    &1.9728\\
            \multirow{2}{*}{\textit{Goldhill}}   &$(2,2)$    & 1.5541    & \textbf{1.7305}  \\
                                                &$(4,4)$    & 1.3051    &1.6224\\
            \multirow{2}{*}{\textit{Baboon}}     &$(2,2)$    & 0.5687    & \textbf{0.6382}  \\
                                                &$(4,4)$    & 0.4163    &0.5587\\
            \bottomrule
        \end{tabular}
    \end{center}
\end{table}

Data embedding capacity is a key metric for evaluating RDH-EI schemes. It is measured by the embedding rate (ER), which quantifies the average number of bits that can be embedded into each pixel of the image, referred to as bits per pixel (\textit{bpp}). 
The high-capacity scheme can vacate large spaces within encrypted images for data embedding due to the correlation in the encryption process. ER of the high-capacity scheme is related to $r$, $n$, $S$ and can be described by Eq.(\ref{eq:ER_of_SchemeI}):
\begin{equation}
    ER = \frac{(BS -1)\times (r-1)\times 8}{BS\times n} bpp
    \label{eq:ER_of_SchemeI}
\end{equation}
Eq.(\ref{eq:ER_of_SchemeI}) can be divided into two components: $\frac{BS - 1}{BS} \times 8$ and $\frac{r - 1}{n}$. For the first component, as $S$ increases, this term approaches 8. For the second component, with $r$ in the range $[2, n]$, increasing $r$ makes this term larger, while increasing $n$ makes it smaller. If $r=n$, then increasing $n$ also increases the second component. Table~\ref{tab:ER1} shows the ERs of the high-capacity scheme under different $r$, $n$ and $S$. When $r=6, n=6$ and $S=8$, the ER reaches its peak value of 6.5625 \textit{bpp}.

In comparison, because encrypted images are smaller and offer fewer blocks for embedding, the size-reduced scheme has a lower embedding rate. Table~\ref{TAB:ER2} presents the ERs of the six test images under different $(r,n)$ and block sizes. ER of the size-reduced scheme depends on $r$, $n$, $S$, and the smoothness of the original images. Increasing $S$ raises the ER while increasing $r$ reduces the size of encrypted images and lowers the ER. Conversely, increasing $n$ enlarges encrypted images and raises the ER.

\section{Comparison and Analysis} \label{sec: Comparison and Analysis}
To illustrate the superiority of our proposed schemes, we conduct comparisons on various aspects. Firstly, we compare the data expansion rate with homomorphic encryption-based and secret sharing-based RDH-EI schemes. Next, 
we assess the ER with some state-of-the-art traditional VRAE-based RDH-EI schemes and secret sharing-based RDH-EI schemes. Then, we evaluate the running time in comparison with secret sharing-based RDH-EI schemes. Finally, we analyze the overall security of our scheme.
\subsection{Data Expansion}
\begin{table}[!t]
  \small
  \begin{center}
      \caption{Data expansion rates of the homomorphic encryption-based and secret sharing-based RDH-EI methods}
      \setlength\tabcolsep{5pt}
      \begin{tabular}{lcc}
        \toprule
        Methods & Content owner & Each data hider\\
        \midrule
        Ke~$et~al.$~\cite{ke_fully_2020}   & 256 & 256     \\
        Li~$et~al.$~\cite{li_histogram_2017}   & 128 & 128     \\
        Chen~$et~al.$~\cite{chen2019new} & $n$ & 1       \\
        Qin~$et~al.$~\cite{qin2021reversible}  & $n$ & 1       \\
        Chen~$et~al.$~\cite{chen_secret_2020} & $n$ & 1       \\  
        Yu~$et~al.$~\cite{yu_reversible_2023}   & $n$ & 1       \\
        Hua~$et~al.$~\cite{hua_reversible_2023}  & $n$ & 1       \\
        ours(capacity)    & $n$ & 1       \\
        ours(size)   & $\approx n-\frac{(BS - 1)\times (r-1)}{BS}$    & $\approx 1-\frac{(BS - 1)\times (r-1)}{BS \times n}$        \\  
        \bottomrule
      \end{tabular}
    \label{tab:data expansion}
    \end{center}
\end{table}
In an RDH-EI scheme, data expansion occurs when the size of the marked encrypted image exceeds that of the original image. This expansion is quantified by the expansion rate, defined as the ratio of the total number of bits in the marked encrypted image to those in the original image. Table~\ref{tab:data expansion} compares the expansion rates of homomorphic encryption-based~\cite{ke_fully_2020,li_histogram_2017} and secret sharing-based RDH-EI schemes~\cite{chen2019new,qin2021reversible,yu_reversible_2023,hua_reversible_2023}. The results show that homomorphic encryption-based methods have significantly higher expansion rates than secret sharing-based schemes. 

In previous secret sharing-based schemes and the high-capacity scheme, the data expansion rate is $ n $ for the content owner and 1 for each data hider. The size-reduced scheme achieves a lower data expansion rate by discarding certain encrypted image pixels, effectively reducing network bandwidth usage. In the size-reduced scheme, increasing $r$ and $S$ reduces data expansion, while increasing $n$ leads to higher expansion. Notably, $r$ must be less than or equal to $n$. When $r$, $S$, and $n$ are small, changes in their values have a significant impact on data expansion.

\subsection{Embedding Capacity}

\subsubsection{Comparison with Traditional VRAE-Based Schemes}

\begin{table}[!t]
\small
\centering
\caption{Average embedding rates of different schemes on different datasets.}
\setlength\tabcolsep{20pt}
\begin{tabular}{ccc}
\toprule
Dataset & Scheme & ER  \\
\midrule
\multirow{5}{*}{BOSSBase}   & Yu~$et~al.$~\cite{yu_reversible_2022}       & 3.2045   \\
                            & Yin~$et~al.$~\cite{yin2020reversible}       & 3.625   \\
                            & Yu~$et~al.$~\cite{yu2021reversible}         & 3.6823 \\
                            & Qiu~$et~al.$~\cite{qiu_high-capacity_2022}  & 3.924  \\
                            & ours(capacity)                         & \textbf{5.9063} \\
                            & ours(size)                          &  2.7798\\  
\multirow{5}{*}{BOWS-2}   & Yu~$et~al.$~\cite{yu_reversible_2022}       & 3.1145   \\
                          & Yin~$et~al.$~\cite{yin2020reversible}       & 3.495   \\
                          & Yu~$et~al.$~\cite{yu2021reversible}         & 3.4568 \\
                          & Qiu~$et~al.$~\cite{qiu_high-capacity_2022}  & 3.793 \\
                          & ours(capacity)                         & \textbf{5.9063} \\
                          & ours(size)                          &  2.6222\\
\bottomrule
\end{tabular}
\label{tab.er compare on datasets}
\end{table}

We compare the ER between our proposed schemes with several state-of-the-art VRAE-based schemes~\cite{yu_reversible_2022,yin2020reversible, yu2021reversible, qiu_high-capacity_2022} on two commonly used datasets, namely BOSSBase and BOWS-2. The results are presented in Table~\ref{tab.er compare on datasets}. The parameter settings follow those of the original papers.
In our schemes, $S$ is set as $8 \times 8$, $r$ and $n$ are set as $4$. The average ERs of the high-capacity scheme on BOSSBase and BOW-2 are both 5.9063~\textit{bpp}, which is the highest and most stable among these schemes. For the size-reduced scheme, its average ERs are 2.7798~\textit{bpp} and 2.6222~\textit{bpp}, respectively. Considering its size-reduced encrypted images, it remains practical and usable. 

\subsubsection{Comparison with Secret Sharing-Based Schemes}

\begin{figure*}[!t]
    \centering
    \begin{subfigure}{0.25\textwidth}
        \includegraphics[width=\linewidth]{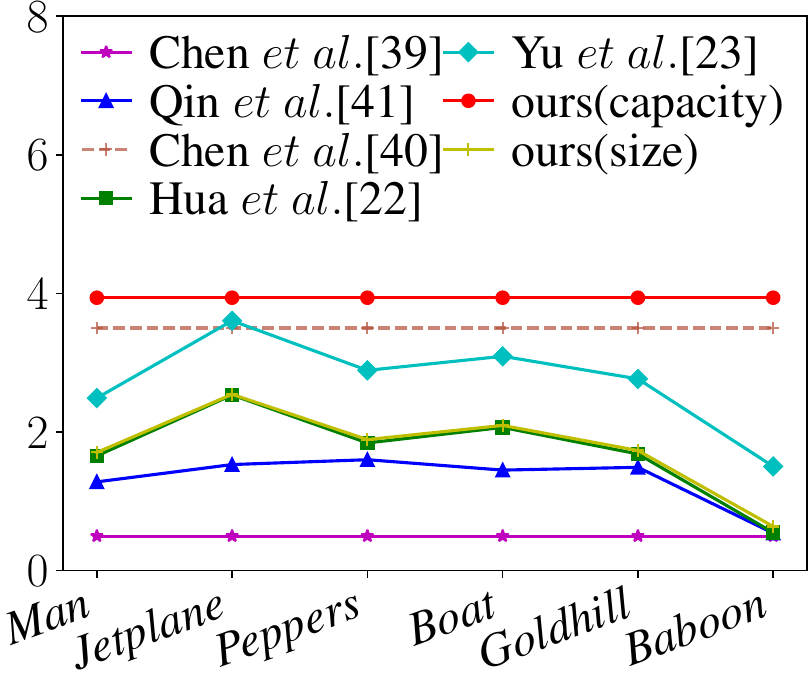}
        \caption{}
    \end{subfigure}
    \hspace{-7pt} 
    \begin{subfigure}{0.25\textwidth}
        \includegraphics[width=\linewidth]{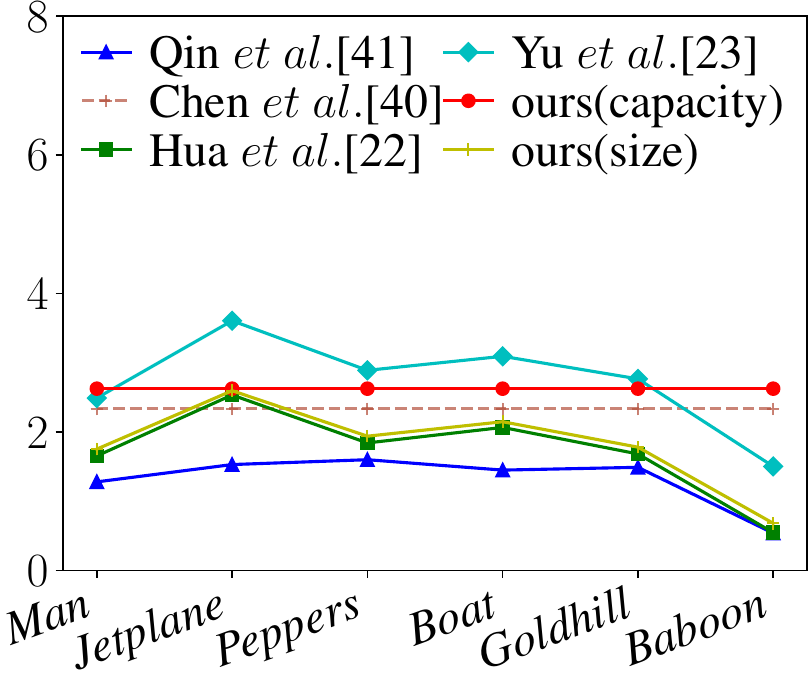}
        \caption{}
    \end{subfigure}
    \hspace{-7pt} 
    \begin{subfigure}{0.25\textwidth}
        \includegraphics[width=\linewidth]{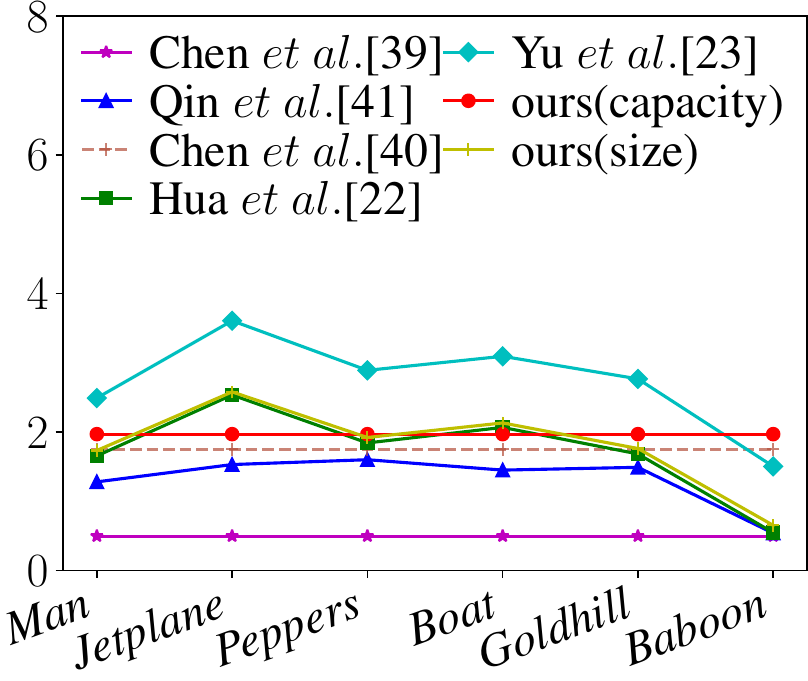}
        \caption{}
    \end{subfigure}
    \hspace{-7pt} 
    \begin{subfigure}{0.25\textwidth}
        \includegraphics[width=\linewidth]{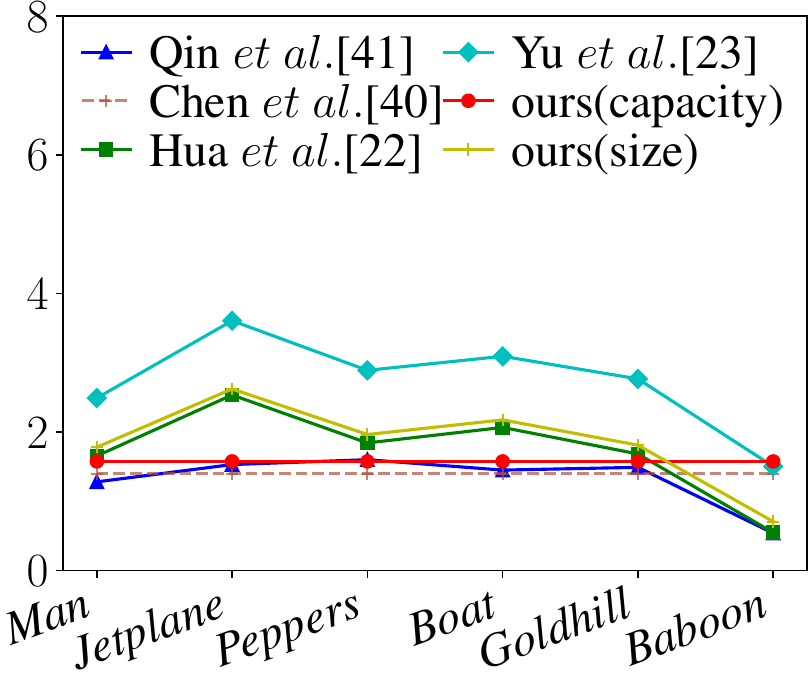}
        \caption{}
    \end{subfigure}

    \begin{subfigure}{0.25\textwidth}
        \includegraphics[width=\linewidth]{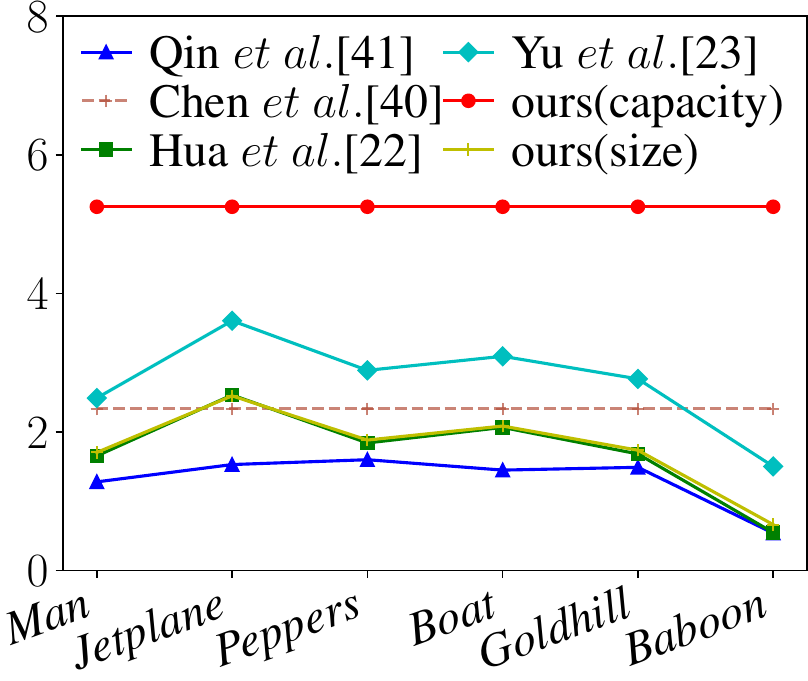}
        \caption{}
    \end{subfigure}
    \hspace{-7pt} 
    \begin{subfigure}{0.25\textwidth}
        \includegraphics[width=\linewidth]{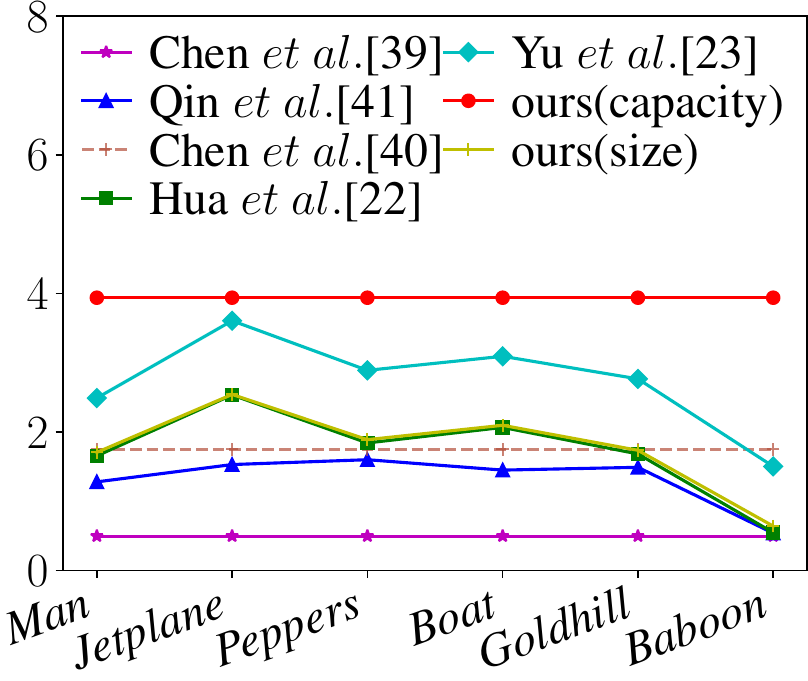}
        \caption{}
    \end{subfigure}
    \hspace{-7pt} 
    \begin{subfigure}{0.25\textwidth}
        \includegraphics[width=\linewidth]{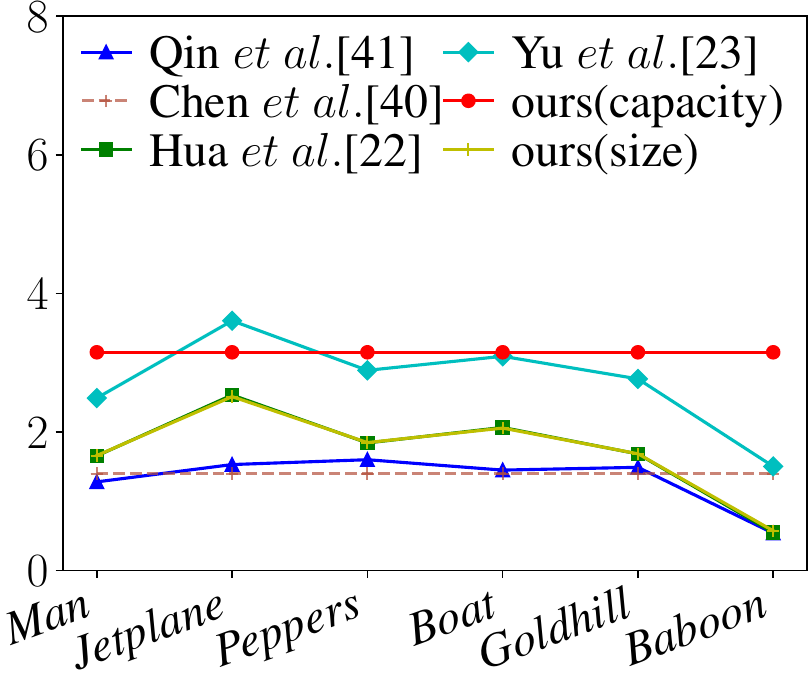}
        \caption{}
    \end{subfigure}
    \hspace{-7pt} 
    \begin{subfigure}{0.25\textwidth}
        \includegraphics[width=\linewidth]{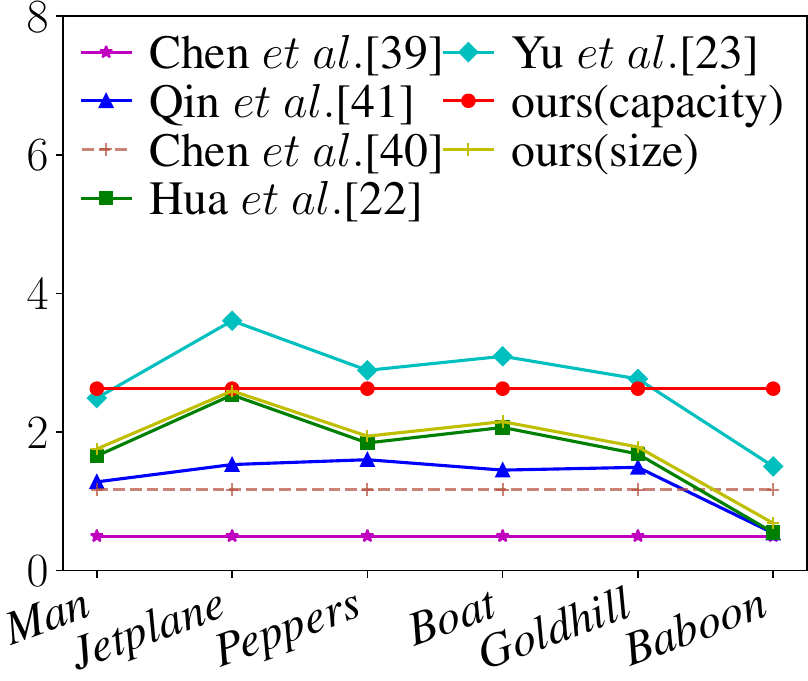}
        \caption{}
    \end{subfigure}

    \begin{subfigure}{0.25\textwidth}
        \includegraphics[width=\linewidth]{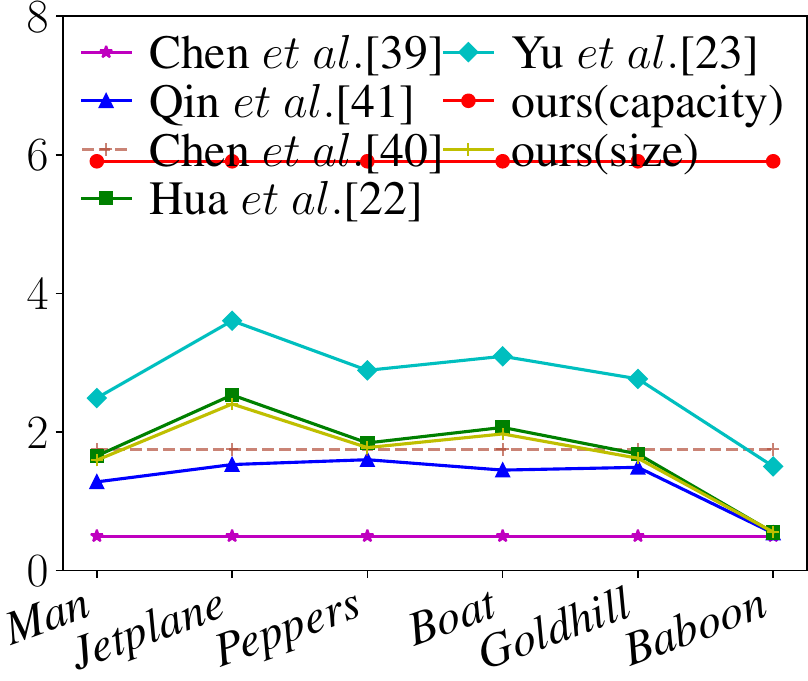}
        \caption{}
    \end{subfigure}
    \hspace{-7pt} 
    \begin{subfigure}{0.25\textwidth}
        \includegraphics[width=\linewidth]{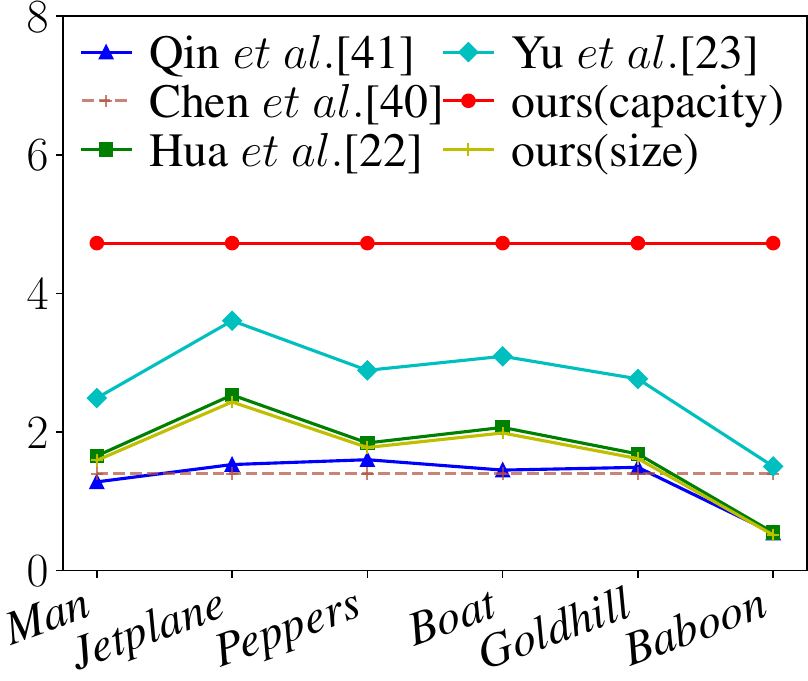}
        \caption{}
    \end{subfigure}
    \hspace{-7pt} 
    \begin{subfigure}{0.25\textwidth}
        \includegraphics[width=\linewidth]{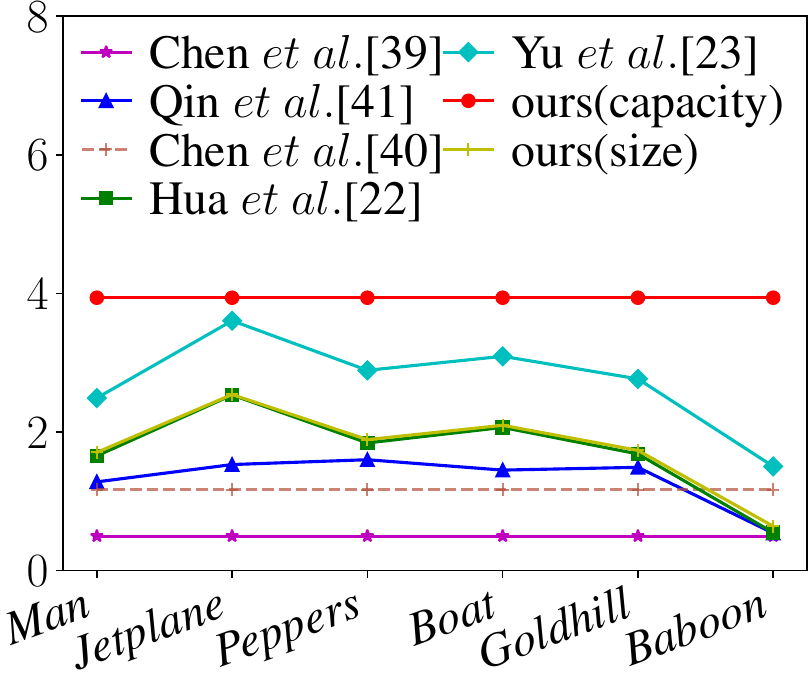}
        \caption{}
    \end{subfigure}
    \hspace{-7pt} 
    \begin{subfigure}{0.25\textwidth}
        \includegraphics[width=\linewidth]{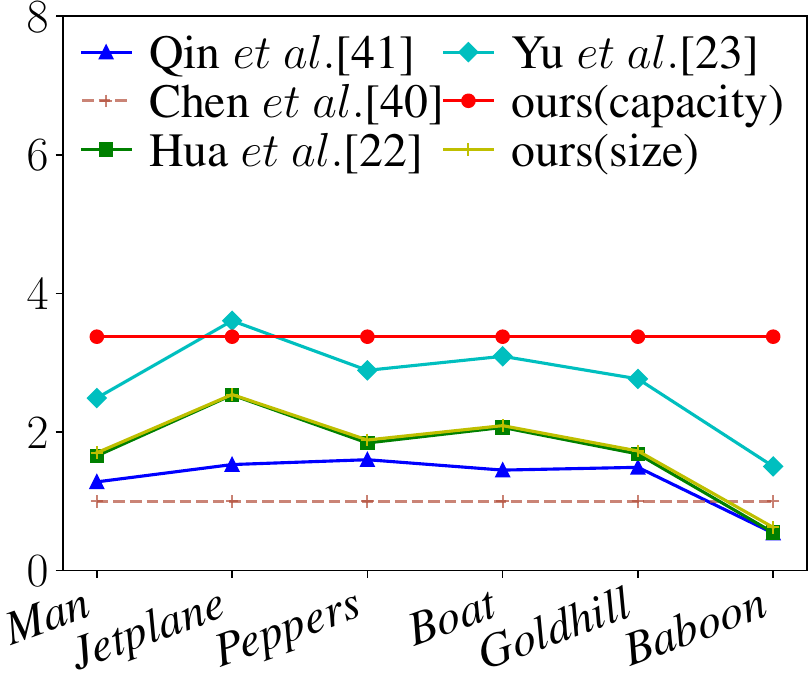}
        \caption{}
    \end{subfigure}

    \begin{subfigure}{0.25\textwidth}
        \includegraphics[width=\linewidth]{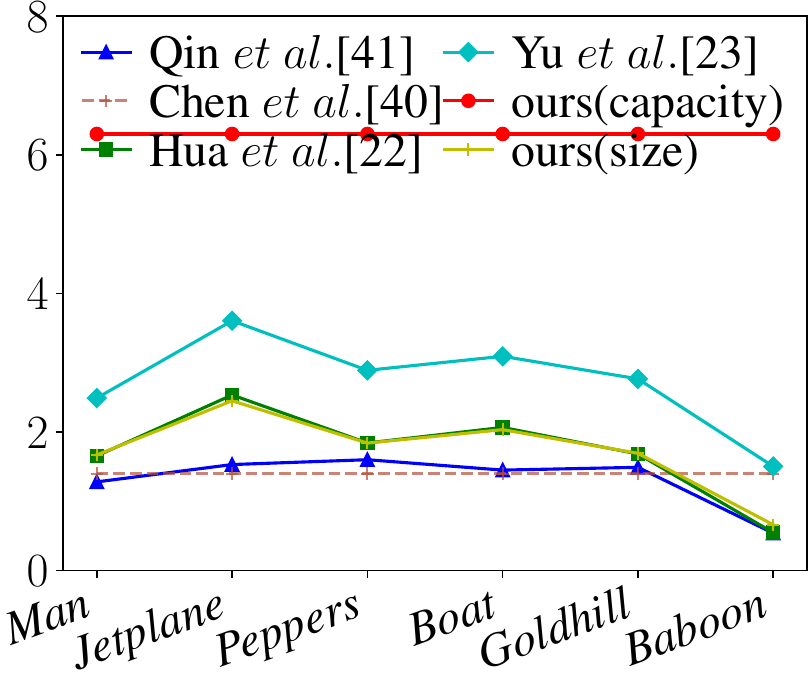}
        \caption{}
    \end{subfigure}
    \hspace{-7pt} 
    \begin{subfigure}{0.25\textwidth}
        \includegraphics[width=\linewidth]{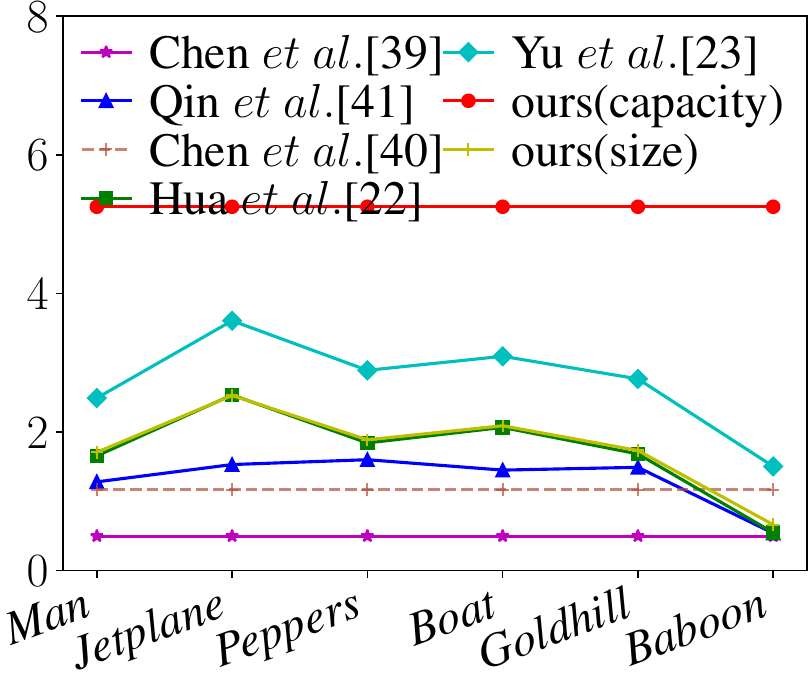}
        \caption{}
    \end{subfigure}
    \hspace{-7pt} 
    \begin{subfigure}{0.25\textwidth}
        \includegraphics[width=\linewidth]{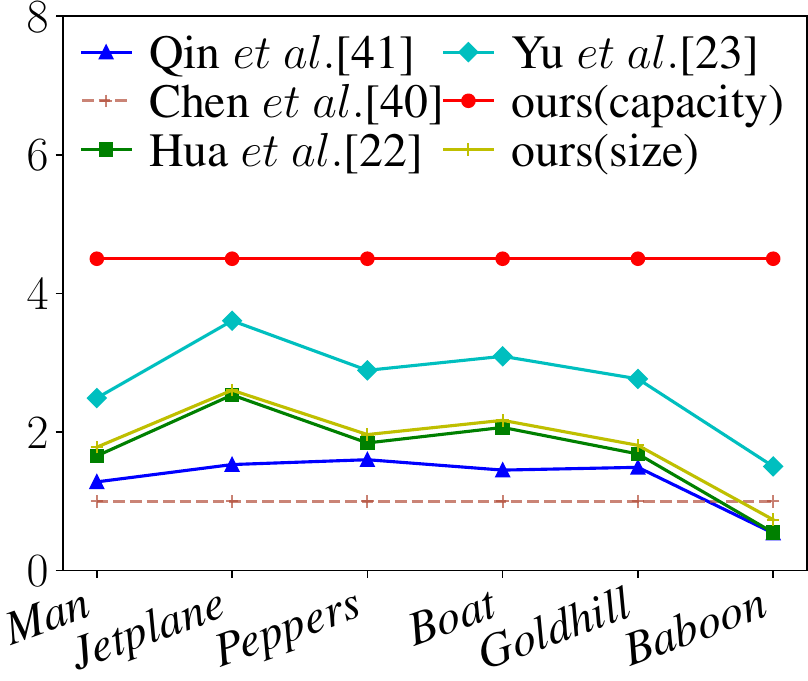}
        \caption{}
    \end{subfigure}
    \hspace{-7pt} 
    \begin{subfigure}{0.25\textwidth}
        \includegraphics[width=\linewidth]{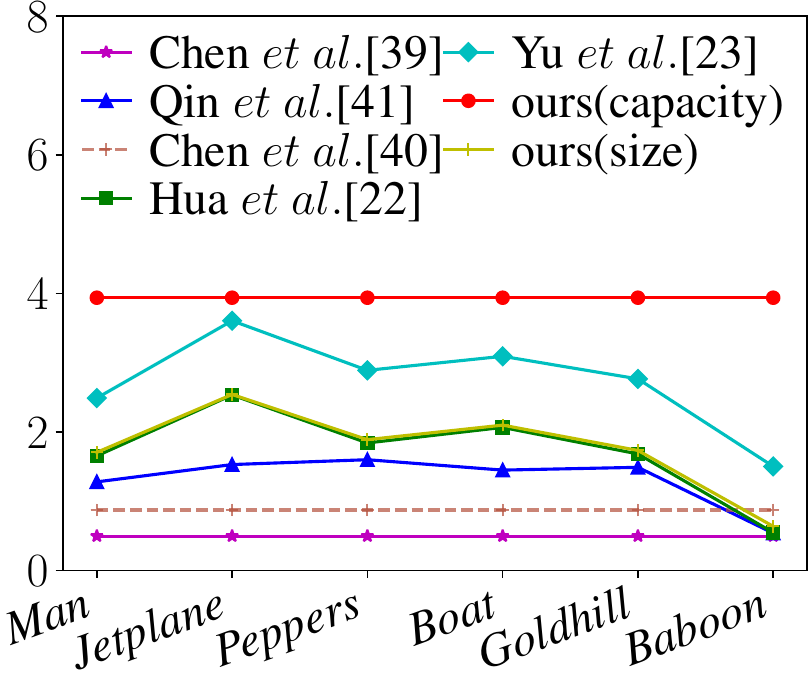}
        \caption{}
    \end{subfigure}
    \caption{Embedding rates of different secret sharing-based RDH-EI schemes with different $(r,n)$-thresholds.}
    \label{Fig:ERs on images}
    \vspace{-10pt}
\end{figure*}

We compare our proposed schemes with existing secret-sharing-based methods~\cite{qin2021reversible, chen_secret_2020, hua_reversible_2023, yu_reversible_2023, chen2019new}. These methods utilize parameters $ r $ and $ n $. In schemes~\cite{qin2021reversible, hua_reversible_2023, yu_reversible_2023, chen2019new}, the ER is independent of $ r $ and $ n $. 
Chen~$et~al.$'s scheme~\cite{chen2019new} cannot be applied when $ r $ is odd. Another scheme~\cite{chen_secret_2020} computes ER based on $ l $ and $ n $, where $ l $ represents the number of replaced bit-planes in each share. For comparison, we set $ l $ to its maximum value of 7. In contrast, the ERs of our schemes depend on $ r $, $ n $, and $ S $. 
For schemes~\cite{hua_reversible_2023, yu_reversible_2023} and our proposed schemes, the block size is fixed at $ 8 \times 8 $. For the scheme of Qin~$et~al.$~\cite{qin2021reversible}, the version with the highest ER is selected. We vary $ r $ from 2 to 5 and $ n $ from $ r $ to $ r+3 $ for all schemes. And other parameters remain consistent with those in the original studies.

Fig.~\ref{Fig:ERs on images} presents the ERs of various secret sharing-based schemes under different $(r,n)$-thresholds. Chen~$et~al.$'s scheme~\cite{chen_secret_2020} achieves an ER of $7/n$, resulting in the highest rate of 3.5~\textit{bpp} when $n=2$. The high-capacity scheme consistently outperforms Chen~$et~al.$'s scheme~\cite{chen_secret_2020} across all images and scenarios. Compared to other approaches, the high-capacity scheme offers superior and more stable embedding capacity under certain $(r,n)$ conditions. The size-reduced scheme, by contrast, demonstrates moderate embedding capability among the evaluated options.

\begin{table*}[!t]
  \small
  \setlength\tabcolsep{4pt}
  \centering
  \caption{Running time comparison (in seconds) of different secret sharing-based schemes for image \textit{Jetplane}.}
  \begin{tabular}{ccccccccc}
    \toprule
    $(r,n)$ & Parties & Chen~$et~al.$~\cite{chen2019new} & Qin~$et~al.$~\cite{qin2021reversible} & Chen~$et~al.$~\cite{chen_secret_2020} & Hua~$et~al.$~\cite{hua_reversible_2023} & Yu~$et~al.$~\cite{yu_reversible_2023} & ours(capacity) & ours(size)\\
    \midrule
    \multirow{3}{*}{(2,2)}      & Content owner & 0.0263 & 0.0096 & 0.0059 & 0.0017 & 0.0289 & 0.0011 & 0.0020 \\
                                & Data hider    & 0.0061 & 0.0064 & 0.0027 & 0.0933 & 0.1231 & 0.0008 & 0.0229 \\
                                & Receiver      & 0.0222 & 0.0476 & 0.0346 & 0.2123 & 0.2827 & 0.0084 & 0.0852 \\
    \multirow{3}{*}{(3,3)}      & Content owner & -      & 0.0212 & 0.0095 & 0.0036 & 0.0391 & 0.0022 & 0.0026 \\
                                & Data hider    & -      & 0.0064 & 0.0030 & 0.0952 & 0.1268 & 0.0010 & 0.0142 \\
                                & Receiver      & -      & 0.0632 & 0.0654 & 0.3057 & 0.4210 & 0.0178 & 0.0874 \\
    \multirow{3}{*}{(4,4)}      & Content owner & 0.0608 & 0.0454 & 0.0113 & 0.0052 & 0.0637 & 0.0038 & 0.0033 \\
                                & Data hider    & 0.0277 & 0.0062 & 0.0033 & 0.0919 & 0.1259 & 0.0010 & 0.0105 \\
                                & Receiver      & 0.0699 & 0.0985 & 0.1037 & 0.3991 & 0.5668 & 0.0383 & 0.1085 \\
    \bottomrule
  \end{tabular}
  \label{TAB:running time}
\end{table*}

\begin{table}[!t]
\small
\centering\caption{Information entropy of the original images as well as encrypted images generated by the high-capacity scheme and the size-reduced scheme.}
\begin{tabular}{ccccc|cc}
\toprule
Image & $\bm{I}$ & $S$ & $\bm{EI}(0)$ & $\bm{EI}(1)$ & $\bm{EI}(0)$ & $\bm{EI}(1)$ \\
\midrule
\multirow{2}{*}{\textit{Man}} &\multirow{2}{*}{7.3574}   & 4 & 7.9984 & 7.9986 & 7.9972 & 7.9975 \\
                                                        && 8 & 7.9977 & 7.9974 & 7.9946 & 7.9944 \\
\multirow{2}{*}{\textit{Jetplane}} &\multirow{2}{*}{6.6776}  & 4  & 7.9983 & 7.9983 & 7.9959 & 7.9964 \\
                                                        && 8 & 7.9957 & 7.9956 & 7.9906 & 7.9905 \\
\multirow{2}{*}{\textit{Peppers}}& \multirow{2}{*}{7.5715} & 4  & 7.9987 & 7.9985 & 7.9971 & 7.9972 \\
                                                        && 8 & 7.9969 & 7.9974 & 7.9955 & 7.9959 \\
\multirow{2}{*}{\textit{Boat}} &\multirow{2}{*}{7.1238} & 4  & 7.9984 & 7.9985 & 7.9969 & 7.9972 \\
                                                        && 8  & 7.9946 & 7.9957 & 7.9950 & 7.9927 \\
\multirow{2}{*}{\textit{Goldhill}}& \multirow{2}{*}{7.4778}& 4  & 7.9985 & 7.9988 & 7.9980 & 7.9981 \\
                                                        && 8 & 7.9980 & 7.9972 & 7.9967 & 7.9969 \\
\multirow{2}{*}{\textit{Baboon}} &\multirow{2}{*}{7.3579} & 4 & 7.9990 & 7.9990 & 7.9985 & 7.9982 \\
                                                        && 8 & 7.9988 & 7.9988 & 7.9980 & 7.9965 \\
\bottomrule
\end{tabular}
\label{TAB:Entropy}
\end{table}

\subsection{Running Time}
We evaluate the running time of our schemes and other secret sharing-based schemes at an ER of 0.4~\textit{bpp} using the test image \textit{Jetplane}. The block size for the schemes of Yu~$et~al.$~\cite{yu_reversible_2023}, Hua~$et~al.$~\cite{hua_reversible_2023} and ours are set as $4 \times 4$. The parameter $l$ is set as 2 in the scheme of Chen~$et~al.$~\cite{chen_secret_2020} to facilitate image recovery at the receiver end. %, all the while upholding the embedding capacity.
Since $r$ and $n$ must be even in the scheme of Chen~$et~al.$~\cite{chen2019new}, we test it under (2,2)-threshold and (4,4)-threshold. 

Table~\ref{TAB:running time} shows the running time results for these schemes. The high-capacity scheme consistently has the shortest running time, as it avoids complex space-vacating methods. The size-reduced scheme also achieves a relatively short running time due to its smaller encrypted images compared to those in Hua~$et~al.$ and Yu~$et~al.$, though it still requires more time than other schemes because of the time-consuming operations needed to create space for embedding.

\subsection{Security Evaluation}
% We evaluate the security of our scheme through both theoretical analysis and experimental validation.

%The results of this comprehensive assessment affirm our scheme's strong security properties.

\subsubsection{Theoretical Analysis}
Our RDH-EI schemes leverage Shamir's secret sharing for encryption, ensuring the security of the original images. As outlined in~\cite{shamir_how_1979}, Shamir's secret sharing provides robust security, preventing unauthorized access to the original image unless the recovery conditions specified in the scheme are met. Even if an attacker gathers $r$ shares, they cannot reconstruct the original image without the encryption key $K_E$. Additionally, the receiver can still recover the original image even if $n-r$ shares are disrupted.

In the encryption phase, all pixels within a block are encrypted using the same parameters, allowing the secret sharing of each block's pixels to be treated as a single unit. Based on the analysis in~\cite{trappe2006introduction}, the probability of recovering the pixels in a block without any prior knowledge of the original image is $1/256$ per block. For an image divided into $BN$ blocks, the probability of recovering the entire image without prior knowledge is $(1/256)^{BN}$, resulting in a brute-force analysis space of $256^{BN}$. However, increasing the block size reduces the number of blocks, which could weaken the security of the image.

Moreover, Shamir's secret sharing is non-deterministic and randomized. Each instance of secret sharing uses randomly selected coefficients, independent of the encryption key $K_E$. As a result, encrypting the same image multiple times with the same $K_E$ produces distinct encrypted images.

% TODO,待修改
\subsubsection{Experimental Validation}

\begin{figure}[!t]
    \centering
    \begin{subfigure}{0.15\textwidth}
        \includegraphics[width=\linewidth]{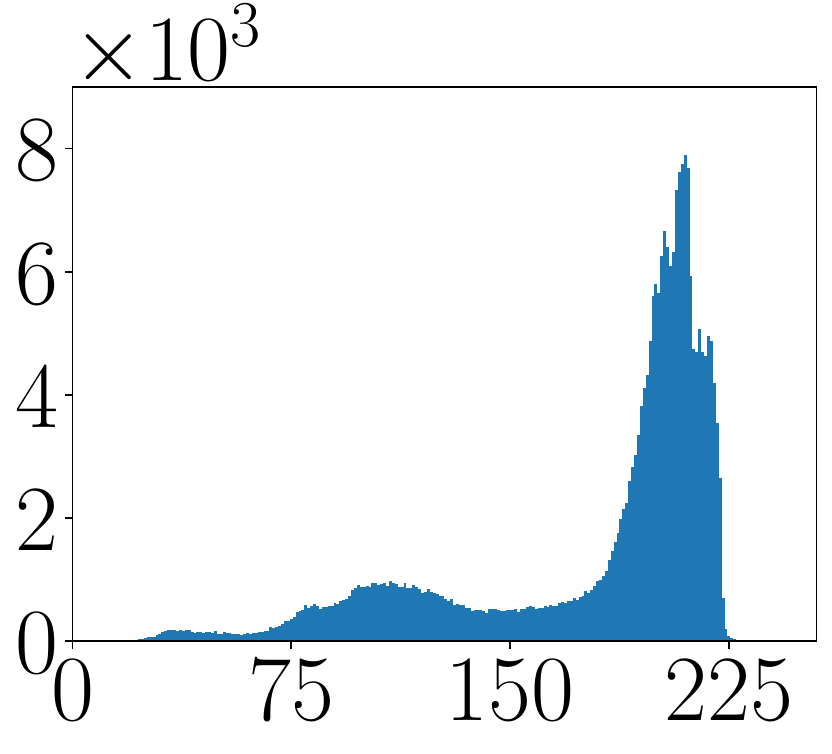}
        \caption{}
    \end{subfigure}
    % \hfill
    \begin{subfigure}{0.15\textwidth}
        \includegraphics[width=\linewidth]{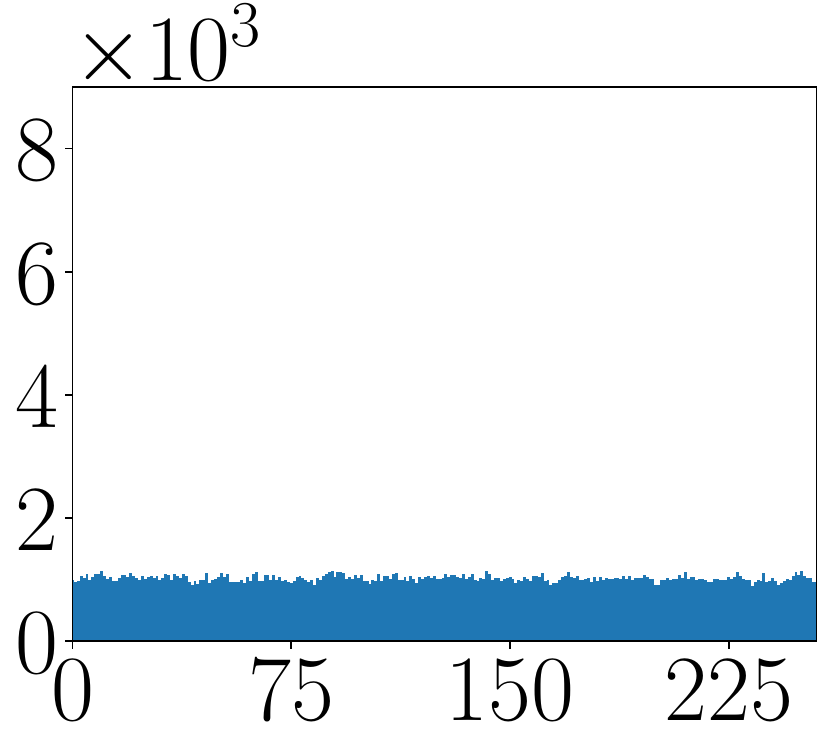}
        \caption{}
    \end{subfigure}
    % \hfill
    \begin{subfigure}{0.15\textwidth}
        \includegraphics[width=\linewidth]{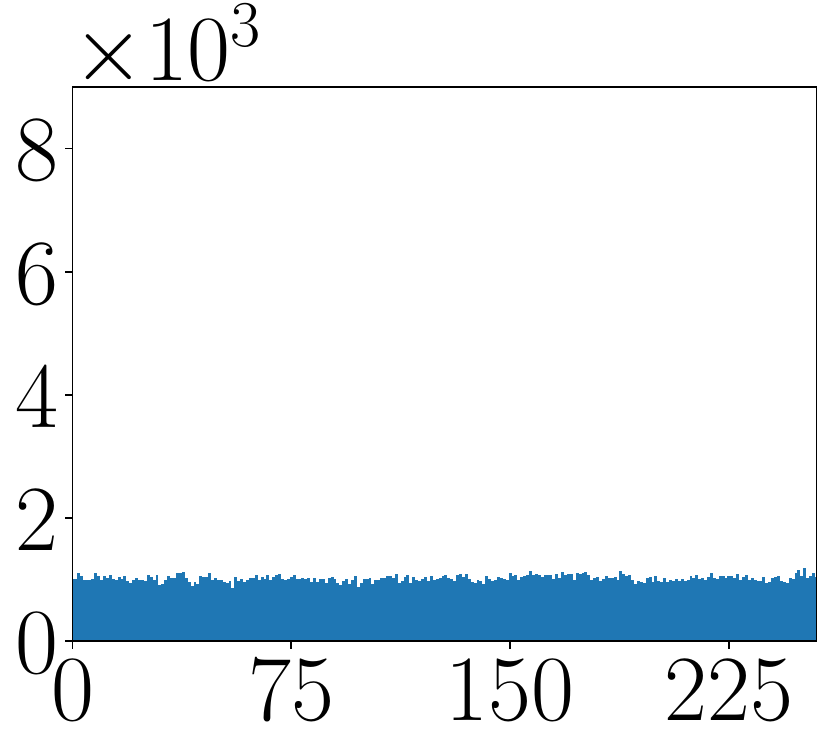}
        \caption{}
    \end{subfigure}
    
    \medskip
    \begin{subfigure}{0.15\textwidth}
        \includegraphics[width=\linewidth]{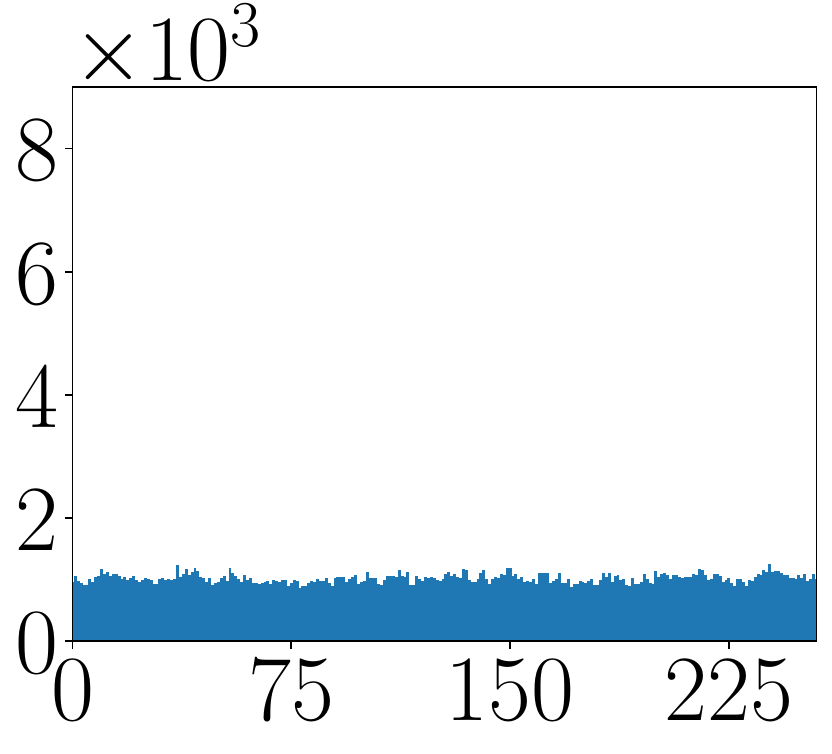}
        \caption{}
    \end{subfigure}
    % \hfill
    \begin{subfigure}{0.15\textwidth}
        \includegraphics[width=\linewidth]{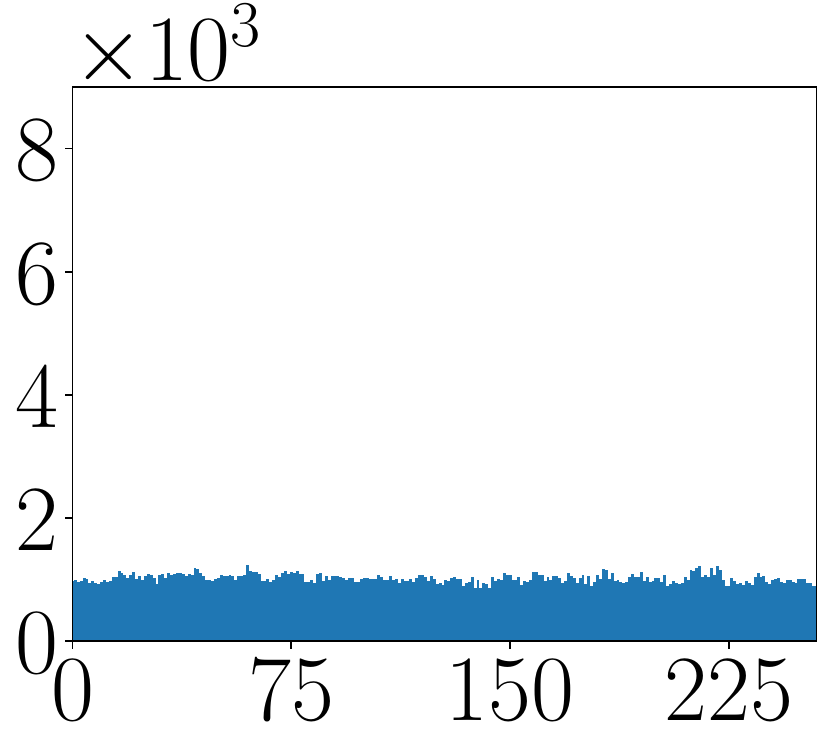}
        \caption{}
    \end{subfigure}
    % \hfill
    \begin{subfigure}{0.15\textwidth}
        \includegraphics[width=\linewidth]{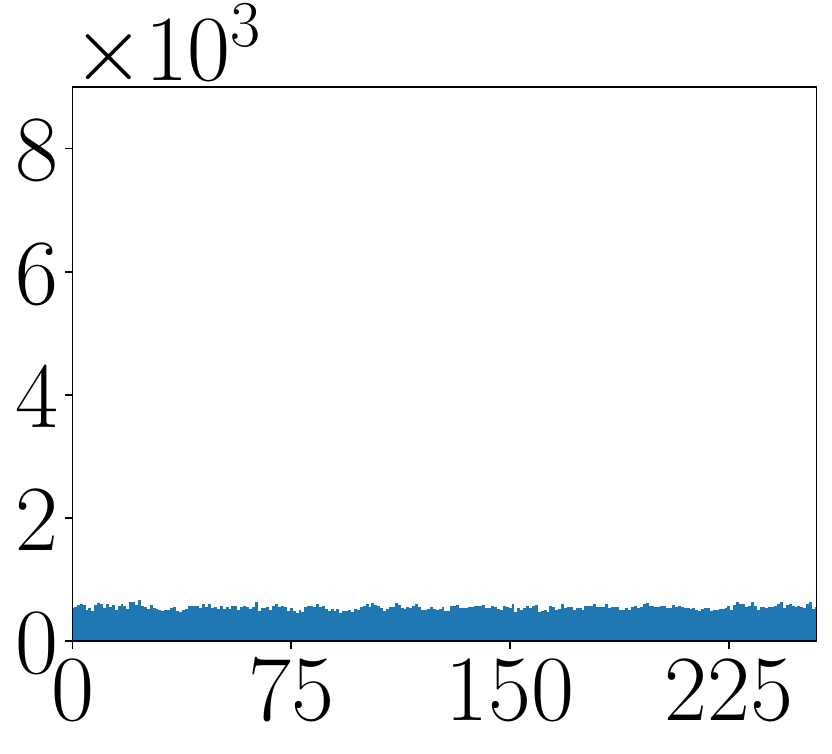}
        \caption{}
    \end{subfigure}
    
    \medskip
        \begin{subfigure}{0.15\textwidth}
        \includegraphics[width=\linewidth]{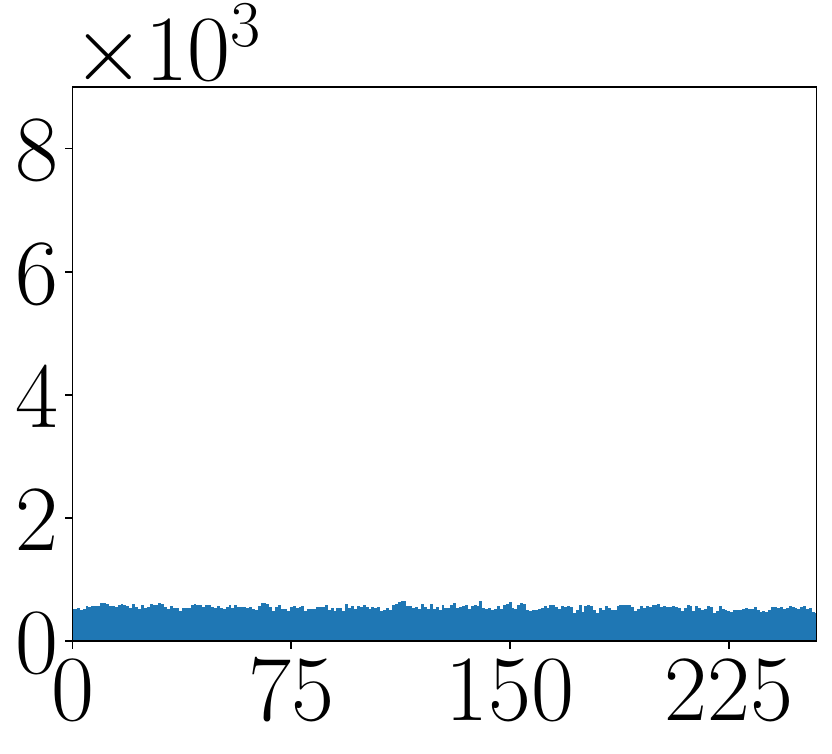}
        \caption{}
    \end{subfigure}
    % \hfill
    \begin{subfigure}{0.15\textwidth}
        \includegraphics[width=\linewidth]{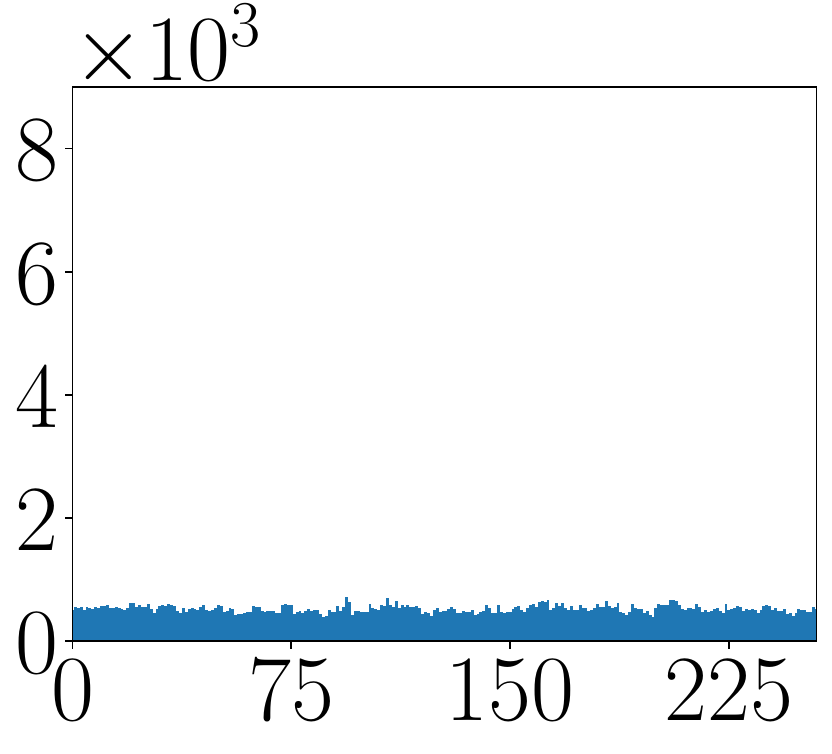}
        \caption{}
    \end{subfigure}
    % \hfill
    \begin{subfigure}{0.15\textwidth}
        \includegraphics[width=\linewidth]{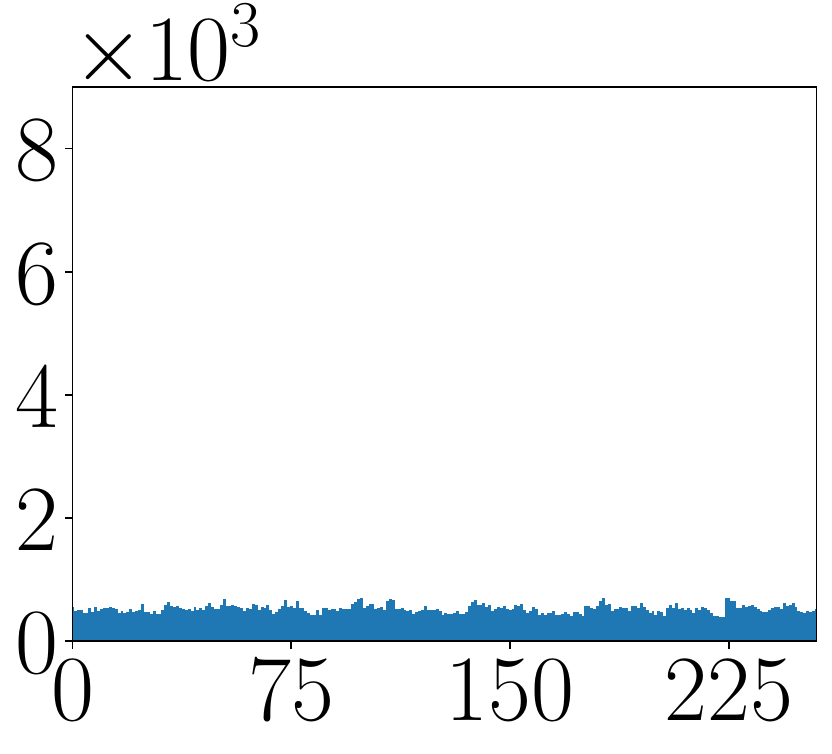}
        \caption{}
    \end{subfigure}

    \caption{Pixel distribution histograms of the image \textit{Jetplane} as well as encrypted images of the high-capacity scheme and the size-reduced scheme. (a) Image \textit{Jetplane}; (b)-(c) the encrypted images of the high-capacity scheme with block size $4\times 4$;(d)-(e) the encrypted images of the high-capacity scheme with block size $8\times 8$;(f)-(g) the encrypted images of the size-reduced scheme with block size $4\times 4$;(h)-(i) the encrypted images of the size-reduced scheme with block size $8\times 8$.}
    \label{Fig:hist of pixels}
    \vspace{-15pt}
\end{figure}

Information entropy quantifies the randomness of images. With the parameter $ p $ set to 256, the ideal entropy value for encrypted images is $ \log_2 256 = 8 $. Table~\ref{TAB:Entropy} presents the entropy values for the original images and the encrypted images generated by the high-capacity scheme and the size-reduced scheme. The results indicate that the entropy values of the encrypted images approach the ideal value. However, a slight decrease in entropy is observed for encrypted images with a block size of $ 8 \times 8 $ compared to those with a block size of $ 4 \times 4 $. 

Fig.~\ref{Fig:hist of pixels} illustrates the histograms of the image \textit{Jetplane} and its encrypted versions generated by our schemes. The histograms of the encrypted images exhibit a near-uniform distribution, reflecting high randomness. Notably, the histogram of the encrypted image with a block size of $ 4 \times 4 $ is slightly more uniform than that of the image with a block size of $ 8 \times 8 $. Additionally, the size-reduced scheme produces encrypted images with lower pixel intensities compared to the high-capacity scheme, resulting in lower histogram peaks.

\section{Conclusion} \label{sec: conclusion}

In this paper, we propose two novel RDH-EI schemes based on secret sharing: the high-capacity scheme and the size-reduced scheme. The high-capacity scheme achieves high embedding capacity by directly generating space for data hiding, while the size-reduced scheme reduces data expansion by producing smaller encrypted images.
We first discuss the intrinsic correlation in the encryption phase when secret sharing is performed based on blocks, which can reduce the number of shares required to reconstruct the original pixel. Building on this, we develop two space-vacating methods and the RDH-EI schemes corresponding to them:
\begin{enumerate}
    \item The high-capacity RDH-EI scheme: The data hider embeds information directly into the encrypted image following a predefined pattern, eliminating the need for complex space-vacating techniques. This approach achieves higher and more stable embedding rates compared to existing methods.
    \item The size-reduced RDH-EI scheme: The content owner generates size-reduced encrypted images by discarding unnecessary shares, significantly reducing data expansion and ensuring efficient bandwidth usage. The data hider then employs traditional space-vacating techniques for embedding.
\end{enumerate}
Experimental results demonstrate that the high-capacity scheme outperforms existing RDH-EI schemes in embedding capacity, while the size-reduced scheme achieves lower data expansion, offering a more efficient alternative. Our schemes provide practical solutions to the challenges of embedding capacity and data expansion in RDH-EI, making them suitable for real-world applications such as medical imaging and cloud storage.

\iffalse
\section*{Acknowledgment}
This work was supported by the National Natural Science Foundation of China under Grants 62071142, 62372125, 62372128, and 62072055, by the Guangdong Basic and Applied Basic Research Foundation under Grants 2023A1515010714, 2023A1515011575, and 2024A1515011475, by the Guangdong Natural Science Funds for Distinguished Young Scholar under Grant 2023B1515020041, and by the Science and Technology Program of Guangzhou under Grant 2024A03J0092.
\fi
%\balance
\bibliographystyle{ieeetr}
\bibliography{ref}

\end{document}